# A quantitative compendium of COVID-19 epidemiology

Yinon M. Bar-On[1*], Ron Sender[1*], Avi I. Flamholz[2], Rob Phillips[3,4], Ron Milo[1]
[1]Weizmann Institute of Science, Rehovot 7610001, Israel
[2]University of California, Berkeley, CA 94720, USA
[3]California Institute of Technology, Pasadena, CA 91125, USA
[4]Chan Zuckerberg Biohub, 499 Illinois Street, SF CA 94158, USA
* Equal contribution

Abstract:

**Accurate numbers are needed to understand and predict viral dynamics. Curation of high-quality literature values for the infectious period duration or household secondary attack rate, for example, is especially pressing currently because these numbers inform decisions about how and when to lockdown or reopen societies. We aim to provide a curated source for the key numbers that help us understand the virus driving our current global crisis. This compendium focuses solely on COVID-19 epidemiology. The numbers reported in summary format are substantiated by annotated references. For each property, we provide a concise definition, description of measurement and inference methods, and associated caveats. We hope this compendium will make essential numbers more accessible and avoid common sources of confusion for the many newcomers to the field such as using the incubation period to denote and quantify the latent period or using the hospitalization duration for the infectiousness period duration. This document will be repeatedly updated and the community is invited to participate in improving it.**

Structure of the document
For each property covered in the current compendium, we present a set of curated studies from the literature best suited for estimating the specific property. For each study we provide an excerpt containing the estimate, to allow full transparency regarding the source of the estimate. Within each excerpt we use bold typeface to guide the readers' attention to the relevant estimate. In addition to presenting the curated set of studies, we also provide a best estimate for each property. Synthesizing estimates provided by all available studies into a singular quantity is challenging because different studies rely on different methodologies for measuring and estimating the property of interest, and thus each has its own associated biases and caveats. In our analysis, we do not attempt to rigorously assess the credibility of each study, but rather provide a best estimate which is the result of our judgement after reading the relevant background material.

**Figure 1: Key epidemiological properties of COVID-19 and their corresponding numerical values.** Further details are provided in the following dedicated subsections. All graphs are schematics and not data. Some values represent population-level median and do not describe inter-individual variability, which is substantial.

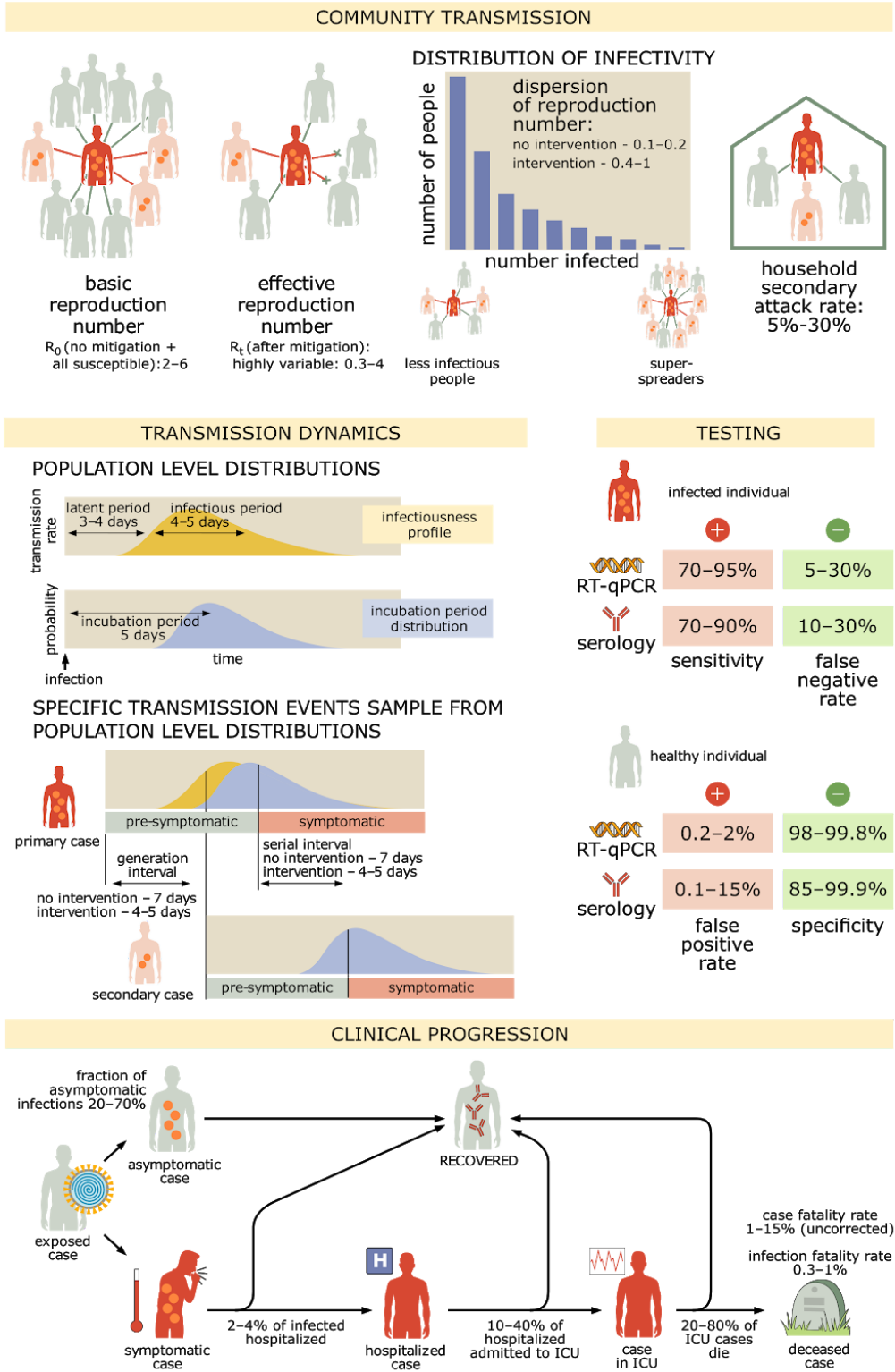

Table of contents:

**Latent period duration**

Definition: Time between infection and the onset of infectiousness. Because this time varies from patient to patient, it is best characterized by a distribution, as shown in the figure.

Best estimate: **3-4 days (population average, inter-individual variation is substantial)**

Methods of measurement and estimation: To calculate the latent period, we would ideally like to know the timing of infection and the time-dependent infectiousness profile for a large number of people. Individual profiles could then be combined to calculate an infectiousness profile (infectiousness as a function of time since infection) for COVID-19. The infectiousness profile is a continuous distribution, so the latent period can then be defined as the time interval that must pass before half-maximal infectiousness. The two main methods of estimating the duration of the latent and infectiousness periods durations are i) use of known infector-infectee pairs, and ii) fitting epidemiological compartmental models.

Estimates:
- Epidemiological models (for early stages in China):
    - (Li et al. 2020) - *"In addition, the median estimates for the latent and infectious periods are approximately **3.69** and 3.48 days, respectively."* Table S1 *"Latency period (Z, days) (95% CIs) (3.48, 3.90)"*
    - (Tian et al. 2020) - Table 3. The table reports an incubation period of **4.9** days (4.32 to 5.47 95% CI). However, the value presented in the table actually represents the latent period and is mistakenly called the incubation period.
- Infector-Infectee pairs
    - 77 pairs from China: (He et al. 2020) - Figure 1C reports the infectiousness profile as a function of days from symptom onset. We use the time it takes the infectiousness to reach half its peak, which happens two days before symptom onset. As symptoms arise after 5 days (see incubation period), this means the latent period is about **3** days.
    - 11 pairs from several countries (Ma et al. 2020) - *" The mean and standard deviation were 7.44 days and 4.39 days for incubation period, **2.52** days and 3.95 days for the upper limit of latent period"*

Caveats:
- Epidemiological models usually use the dynamics of the epidemic to infer a population-level characteristic (e.g. median or mean). The structure of the model inherently determines the distribution of the latent period. For example, in the standard SEIR model, the latent period is assumed to be exponentially distributed.
- Simple SEIR-type models assume infectiousness onset occurs at a specific time, before which a person is not infectious and after which they become fully infectious. A more realistic scenario is that infectiousness increases gradually throughout the infection progression, and thus the definition of the latent period is harder to nail down. We use the time until half maximal infectiousness as the definition for the latent period in infector-infectee pair data.
- When estimating the population median latent period from infector-infectee pairs, there is an inherent assumption that the infectiousness profile of different people is similar. Even though this assumption does not fully capture the variability between individuals for example based on their age or based on the state of their immune system, it is a conscious choice made to reduce the complexity of the models, and results in an effective value that averages across this variability.

- There is often confusion in the literature between the latent period and the incubation period, the latent period being the time before the onset of infectiousness and the incubation period being the time before onset of symptoms. Many manuscripts refer to the properties of the latent period but write it as the incubation period or use the value of the incubation period to describe the latent period.

**Infectious period duration**

Definition: Characteristic duration over which a person is infectious.

Best estimate: **4-5 days (population average, inter-individual variation is substantial)**

Methods of measurement and estimation: The two main methods for estimating the duration of infectiousness is (i) use of data from infector-infectee pairs, and (ii) fitting epidemiological compartmental models. When using data reported by infector-infectee pairs, it is difficult to determine the exact time of infection. Instead, patient-reported timing of symptom onset is used to determine the lag between infector symptoms and infectee symptoms, also known as the serial interval (see "Serial interval/Generation interval" section). This data is combined with independent data on the distribution of incubation periods (the time from infection to symptom onset), to infer a distribution of generation intervals (time from infection of the infector to the infection of the infectee). As it quantifies the probability of transmission per day post infection, the distribution of generation intervals is also commonly used to infer an "infectiousness profile" for infected patients. The difference between the generation interval distribution and the infectiousness profile is that the first represents a distribution of the probability of a secondary infection after a given time interval since the first infection, while the other represents a distribution of those transmission rates. Therefore, the integral of the generation interval distribution is by definition equal to one, while the integral of the infectiousness profile is the reproduction number $R_t$. Thus, to convert the distribution of the generation intervals into an infectiousness profile, it is scaled by the average reproduction number (Ferretti et al. 2020). As this is a continuous distribution, there isn't a single "infectious period" value, but one can define the interval between half the maximal infectiousness as a measure indicative of the infectious period duration. Alternatively, the infectious period can be inferred by fitting epidemiological compartmental models as these models relate the (unknown) infectious period to reported counts of cases and deaths.

Estimates:

- Infector-Infectee pairs
  - (77 pairs from China): (He et al. 2020) - Figure 1C reports the infectiousness profile as a function of days from symptom onset. We quantify the interval between the time points of half the maximal infectiousness based on the infectiousness profile, which is about **4** days.
  - (44 pairs from various countries): (Ferreti et al. 2020) - "*The distribution is best described by a Weibull distribution (AIC=148.4, versus 149.9 for gamma and 152.3 for lognormal distribution) with mean and median equal to **5.0** days and standard deviation of 1.9 days*". The infectiousness profile has the same distribution as the generation time distribution.
- Epidemiological compartmental models (for early stages of the epidemic in China):
  - (Li et al. 2020) - "*In addition, the median estimates for the latent and infectious periods are approximately 3.69 and **3.48** days, respectively.*" Table S1 "*Infectious period for documented infections (Dr , days)  (95% CIs) (3.26, 3.67)* "
  - (Tian et al. 2020) - Table 3. The table reports **5.1** days long infectious period (4.51 to 5.86 95% CI)

Caveats:
- Interventions will affect the duration of the infectious period. For example, isolation of infected individuals with symptoms limits their chances to infect non-household members. This reduces the overall number of infections, and shortens the time in which such infections occur, and hence the infectious period duration.
- Epidemiological compartmental models usually use the dynamics of the epidemic to infer the population-level characteristic (e.g. median or mean). The structure of the model inherently determines the distribution shape for the infectious period. For example, in the simple Susceptible-Exposed-Infectious-Recovered (SEIR) model, the distribution of the infectious period is assumed to be exponential.
- Simple SEIR-type models assume infectiousness is constant throughout the infectious period, and that the start and end of infectiousness occurs abruptly. A more realistic scenario is that infectiousness increases gradually throughout the progression of the infection and then decreases. In this scenario, the semantics of the infectious period is harder to define. We use the duration in which a person has infectiousness higher than half of the maximal infectiousness as the definition for inference of the infectiousness period from infector-infectee pairs data.
- When estimating the population average infectious period from infector-infectee pairs, there is an inherent assumption that the infectiousness profile of different people is similar. Even though this assumption does not fully capture the variability between individuals for example based on their age or based on the state of their immune system, it is a conscious choice made to reduce the complexity of the models, and results in an effective value that averages across this variability.
- One should be careful not to confuse the infectious period duration with the time from infection until someone can be infected. There is a latent period duration that adds to the infectious period and together these two distributions determine the time from being infected until infecting others. In addition, there is also wide variation among cases in their infectiousness profile, so the infectious period duration should be used as a statistical value rather than the maximal date after which people stop being contagious.
- Another source of confusion is usage of the duration of illness or hospitalization as the duration of infectious period. The infectious period is much shorter than the overall duration of illness (which is often on the order of a few weeks). People's illness is usually prolonged due to the inflammatory response of the immune system. Thus people usually stop being strongly contagious well before they recover from the disease. The dynamics of the epidemic is governed by the infectious period and not by the time of overall illness duration.

**Incubation period duration**

Definition: Time from infection until onset of symptoms in symptomatic cases.

Best estimate: **5-6 days (population average, inter-individual variation is substantial)**

Methods of measurement and estimation: There are two main methods for estimating the incubation period of COVID-19. The first is based on investigation of possible exposure times for reported cases. This usually entails collecting information on possible exposure time for a set of cases, and combining this information with records of the onset of symptoms. A statistical model is used to fit the data with a possible distribution of incubation times. The second method is based on epidemiological models. In this approach, exponential growth is used to model the dynamics of the epidemic, and a certain functional form is assumed for the distribution of the incubation period. The observed number of cases is a combined result of viral transmission (the output of the exponential growth model), and the distribution of incubation period (causing an infected person to show symptoms and thus be detected as a case). The model parameters, both for the exponential growth model, (similar to those described in the effective reproduction number section) as well as for the incubation period, are inferred such that the model's projections match case data as closely as possible.

Estimates:
- Investigation of the possible time for reported cases
    - 181 cases outside Hubei: (Lauer et al. 2020) - "*The median incubation period was estimated to be **5.1** days (95% CI, 4.5 to 5.8 days), and 97.5% of those who develop symptoms will do so within 11.5 days (CI, 8.2 to 15.6 days) of infection. These estimates imply that, under conservative assumptions, 101 out of every 10 000 cases (99th percentile, 482) will develop symptoms after 14 days of active monitoring or quarantine.*"
    - 10 cases in Wuhan: (Li et al. 2020) - "*The mean incubation period was **5.2** days (95% confidence interval [CI], 4.1 to 7.0), with the 95th percentile of the distribution at 12.5 days*".
    - 88 cases of travellers from Wuhan (Backer et al. 2020) - "*The Weibull distribution provided the best fit to the data (Table 1). The mean incubation period was estimated to be **6.4** days (95% credible interval (CI): 5.6–7.7).*"
    - 158 residents from and outside Wuhan (Linton et al. 2020) - "*The mean incubation period was estimated at **5.0** days (95% credible interval [CI]: 4.2, 6.0) when excluding Wuhan residents (n = 52) and **5.6** days (95% CI: 5.0, 6.3) when including Wuhan residents (n = 158).*"
    - 22 patients from South Korea (Ki et al. 2020) - "*The average incubation period was **3.9** days (median, 3.0)*"
    - 50 global cases (Jiang et al. 2020) - "*The corresponding mean and 95% CI were: SARS-CoV-2, **4.9** (4.4-5.5) days*"
    - 291 cases in China (Guan et al. 2020) - "*The median incubation period was **4** days (interquartile range, 2 to 7)*"
    - 49 cases from China (Zhang et al. 2020) - "*We analyzed the time interval from exposure to illness onset for 49 cases (with no travel history to/from Wuhan/Hubei ) identified by prospective contact tracing in 37 clusters. We estimated a mean incubation period of **5.2** days (95%CI: 1.8-12.4), with the 95th percentile of the distribution at 10.5 days*"

- - 93 cases from Guizhou, China [(Ping et al. 2020)](#) - "*The median incubation period was **8.06** days (95% confidence interval [CI]: 6.89 – 9.36).*"
  - 91 cases from Zhejiang, China [(Qian et al. 2020)](#) - "*The median of incubation period was **6** (IQR, 3-8) days*"
  - 262 cases from Beijing, China [(Tian et al. 2020)](#) - "*The median time from contact symptomatic case to illness onset, which is called the incubation period, was **6.7** days*"
  - 22 paired cases from Taiwan and mainland China [(Cheng et al. 2020)](#) - "*Using the data on the 22 paired cases, we estimated that the median incubation period was **4.1** days (95% credible interval [CrI], 0.4-15.8)*"
  - 587 cases from seven countries [(Ma et al. 2020)](#) - "*By pooling individual data from seven countries, we estimated the mean incubation period of COVID-19 to be **7.44** days*"
- Epidemiological models (for early stages in China): [(Li et al. 2020)](#) - "*The mean incubation period was **5.2** days (95% confidence interval [CI], 4.1 to 7.0), with the 95th percentile of the distribution at 12.5 days.*"

<u>Caveats:</u>
- There is often confusion in the literature between the latent period and the incubation period, the latent period is defined as the time before the onset of infectiousness and the incubation period is defined as the time before onset of symptoms. Many manuscripts refer to the properties of the latent period but incorrectly identify it as the incubation period or use the value of the incubation period to describe the latent period.

**Serial interval / Generation interval**

Definition
Generation interval ≡ Average time between infection events in an infector-infectee pair.
Serial interval ≡ Average time between symptom onsets in an infector-infectee pair.
See Figure 1 for a graphical representation.

The serial interval distribution is a combination of the generation interval distribution with the incubation period distribution. The expected values of the serial and generation interval distributions are expected to be equal but their variances are different. The variance of the serial interval equals the sum of variances of the generation interval and the incubation period, and thus is larger. The generation interval is more useful for describing viral spread than the serial interval, because it directly relates to the timing of infection. Unfortunately, this distribution is harder to estimate than the distribution of serial intervals because people do not typically know when they were infected.

Interventions such as isolation of symptomatic patients and contact tracing could shorten these two intervals by replacing lower frequency and shorter duration contact (e.g. at work and social gatherings) with high-frequency and long duration contact at home. The interventions lower the mean and median serial interval by censoring the distribution of serial intervals (among the various infected-infectee couples).

Best estimate:
No interventions: **7 days**
Isolation of symptomatic patients and contact tracing: **4-5 days**

Methods of measurement and estimation:
Estimation of the serial interval is based on identification and tracking of chains of transmission. Identifying the precise moment of infection is challenging, so most researchers use patient-reported timing of symptom onset, which is available in more cases. Data regarding symptom onset of identified pairs of infector-infectee is collected through detailed interviews. The delay between the onset of symptoms of each pair is sometimes used to calculate descriptive statistics (median, mean, standard deviation) and, in other cases, is used to fit an a priori distributional model.
In order to estimate the generation interval from the serial interval, one must account for the incubation period because the infector is not infectious during this time. In some sources, symptom onset timing from infector-infectee pairs is combined with a reported distribution of incubation periods to estimate the generation time by fitting to a parametric distribution. In most cases, the fit distributional parameters are allowed to assume a wide range of values, such that the assumptions have a minimal effect on the fit.

Estimates:
- Starting of the epidemic :
    - Uncontrolled in Wuhan (china)
        - 6 transmission pairs (Li et al. 2020) - "*we estimated that the serial interval distribution had a mean (±SD) of **7.5±3.4 days (95% CI, 5.3 to 19)**"*
        - 43 transmission pairs (Wu et al. 2020b) - "*The mean serial interval is **7.0** (5.8–8.1) days, with a standard deviation of 4.5 (3.5–5.5) days*".

- Vo, Italy - (Lavezzo et al. 2020) - *"From the inferred transmission pairs, we estimated a serial interval distribution with mean **6.90** days (95% CI 2.56-13.39) before lockdown (Figure S7) and 10.12 days (95% CI 1.67-25.90) after the lockdown"*

- Serial interval measured in countries that implemented isolation of symptomatic patients, physical distancing and contact tracing:
    - Korea - (Ki et al. 2020) - *"The mean and median serial interval was estimated to be **6.6** days (range 3-15) and **4.0** days, respectively, based on data of both first-generation and second-generation patients"* (The mean duration between symptom-onset and quarantine/isolation was 4.3 days)
    - Hong Kong
        - 21 transmission chains (S. Zhao, Gao, et al. 2020) - *"Assuming a Gamma distributed model, we estimated the mean of SI at **4.4** days (95%CI: 2.9–6.7) and SD of SI at 3.0 days (95%CI: 1.8–5.8) by using the information of all 21 transmission chains in Hong Kong."*
        - 171 transmission pairs with 94 unique infectors (Adam et al. 2020) - *"Figure 2A shows the empirical serial distribution between all infector-infectee pairs and fitted normal and lognormal distributions. The median serial interval was **4** days (IQR 3 – 8 days), and the mean of the fitted normal distribution was **5.6** days (standard deviation 4.3 days)."*
    - Singapore and Tianjin (China) - (Ganyani et al. 2020) - *"The mean generation time is estimated to be **5.2** days (95%CI, 3.78 - 6.78) for the Singapore data, and **3.95** days (95%CI, 3.01 - 4.91) for the Tianjin data".*
    - Shenzhen - 48 transmission pairs (Bi et al. 2020) - *"we estimate that the serial interval is gamma distributed with mean **6.3** days (95% CI 5.2,7.6) and a standard deviation of 4.2 days (95% CI, 3.1,5.3)".*
    - Outside of Hubei province - 468 transmission events (Zhanwei Du, Xiaoke Xu, et al. 2020) - *"COVID-19 serial intervals seem to resemble a normal distribution more than the commonly assumed gamma or Weibull distributions , which are limited to positive values (Appendix). We estimate a mean serial interval for COVID-19 of **3.96** (95% CI 3.53–4.39) days, with an SD of 4.75 (95% CI 4.46–5.07) days".*
    - Taiwan and mainland China - 22 transmission pairs (Cheng et al. 2020) - *"we estimated that the median incubation period was 4.1 days (95% credible interval [CrI], 0.4-15.8), and the median serial interval was **4.1** days (95% CrI, 0.1-27.8)... A secondary case was excluded from the paired data if the beginning of exposure was after symptom onset of the secondary case".* They also separate the results between the various regions (medians: Taiwan=4.1, mainland China=5, other =2.9) and setting (medians: household=5.2, community:2.9, other=3.3) see supplementary.

- Measurement from multiple countries data aggregating effects of their different interventions:
    - China, Japan, Singapore, South Korea, Vietnam, Germany and Malaysia - 689 transmission pairs (Ma et al. 2020) *"The serial interval was estimated for 689 pairs of infector-infectee and ranged from −5 to 24 days (Table 2), with a median of 6 days. It was smaller than 0 (mean: −2.63 days) for 27 (3.92%) pairs, meaning that the infectee showed symptoms earlier than the infector. Assuming a normal distribution (Figure*

*2(C)), the mean was **6.70** days (95% CI: 6.31, 7.10) and standard deviation 5.20 days (95% CI: 4.91, 5.46)."*
- China, Vietnam, Japan, Malaysia, Hong Kong, Taiwan - 77 transmission pairs (He et al. 2020) *"serial interval was estimated to have a mean of **5.8** days (95% confidence interval (CI), 4.8–6.8 days) and a median of **5.2** days (95% CI, 4.1–6.4 days) based on a fitted gamma distribution, with 7.6% negative serial intervals"*
- Vietnam, South Korea, Germany, Taiwan, Singapore, China - 28 transmission pairs (Nishiura, Linton, and Akhmetzhanov 2020) - *"lognormal distribution was selected as the best-fit model (WAIC = 224.0), while no significant differences from other models were identified. The median serial interval was estimated at **4.0** days (95% credible interval [CrI]: 3.1, 4.9)"*
- China, Vietnam, South Korea, Germany, Italy, Singapore, Hong Kong, Taiwan - 40 transmission pairs (Ferreti et al. 2020) - *"we directly estimated the generation time distribution from 40 source-recipient pairs...The distribution is best described by a Weibull distribution with mean and median equal to **5.0** days and standard deviation of 1.9 days"*.

<u>Caveats</u>:
- The serial interval is affected by policies of isolation and physical distancing. As such, measurements from one place and time (e.g. Wuhan before lockdown) should not be applied to another location (e.g. Lombardy after lockdown) without explicit explanation. As mentioned for example in (Bi et al. 2020): *"It should be noted this estimate includes the effect of isolation on truncating the serial interval. Stratified results show that if the infector was isolated less than 3 days after infection the average serial interval was 3.6 days, increasing to 8.1 days if the infector was isolated on the third day after symptom onset or later"*.
- In general it is hard to track large networks of transmissions. Some sources use very small numbers of infector-infectee pairs (e.g. < 10) to arrive at an estimate.
- Possible biases in the selection of the pairs used for the analysis. In addition, recent contacts are easier to remember than distant contacts leading to bias in the backward contact tracing, toward shorter serial intervals.
- Clusters of infections could make it difficult to determine which of several individuals is the infectee.
- The most common approach to estimate the serial or generation intervals is to use contact-tracing data. In this approach infected individuals are questioned to identify their infectors, and the duration between the infection times of infector and infectee is estimated. This approach is inherently backward-looking, starting from infected individuals and going back in time to identify their infectors. This backward-looking perspective biases the estimate of the serial and generation intervals in the early stages of an epidemic, when cases grow exponentially (Park et al. 2020). Because the number of infected people grows exponentially, infections with short generation or serial intervals will be overrepresented in the sampled population, which will consequently lead to underestimating the generation and serial intervals.

Comparison to other pathogens:
- No lockdown:
    - SARS - (Lipsitch et al. 2003) - *"The mean serial interval in Singapore was **8.4** days (SD = 3.8)"*
    - MERS - (Cauchemez et al. 2016) - *"We estimate that the serial interval (delay between symptom onset in a case and symptom onset in the persons they infect) of MERS-CoV has a mean of **6.8** (95% CI: 6.0, 7.8) days and a SD of 4.1 (95% CI: 3.4, 5.0) days (Fig. 2C)."*
    - Influenza - (Cowling et al. 2009) - *"We estimated the household serial interval of influenza to be between **3** and **4** days"*.
    - Tuberculosis - (Ma et al. 2018) - *"the serial interval estimates from four studies were: **0.57, 1.42, 1.44** and **1.65** years;"*

**Doubling time and exponential growth rate**

Definition: The doubling time is the time it takes the cumulative number of cases to double. Exponential growth rate is the slope of the curve of the log of infected cases versus time. The growth rate r and doubling time τ are connected by the equation: $r = \frac{ln2}{\tau}$ (as $e^{r\tau} = e^{ln2} = 2$).

Best estimate: situation specific (depends on location and interventions)

Methods of measurement and estimation:
The exponential growth rate is usually estimated by an exponential fit to the curve of cumulative infections.

Estimate:
- Outbreak in Wuhan and mainland China:
  - Estimates on the basis of the reported number of cases:
    - (Muniz-Rodriguez et al. 2020a) - *"During January 20–February 9, the harmonic mean of the arithmetic means of the doubling times estimated from cumulative incidence ranged from **1.4** (95% CI 1.2–2.0) days in Hunan Province to **3.1** (95% CI 2.1–4.8) days in Xinjiang Province. We estimated doubling time as **2.5** (95% CI 2.4–2.6) days in Hubei Province. The cumulative incidence doubled 6 times in Hubei Province during the study period. The harmonic mean of the arithmetic means of doubling times for mainland China except Hubei Province was **1.8** (95% CI 1.5–2.3) days. Fujian, Guangxi, Hebei, Heilongjiang, Henan, Hunan, Jiangxi, Shandong, Sichuan, and Zhejiang provinces had a harmonic mean of the arithmetic means of doubling times <2 days (Figure; Appendix Figure 1)."*
    - (Li et al. 2020) - *"In the epidemic curve up to January 4, 2020, the epidemic growth rate was 0.10 per day (95% CI, 0.050 to 0.16) and the doubling time was **7.4** days (95% CI, 4.2 to 14)."*
  - Estimates on the basis of the number of cases exported from Wuhan to cities outside in mainland China and other countries:
    - (Zhanwei Du et al. 2020) - *"By fitting our epidemiologic model to data on the first 19 cases reported outside of China, we estimate an epidemic doubling time of **7.31** days (95% CrI 6.26–9.66 days) and a cumulative total of 12,400 (95% CrI 3,112–58,465) infections in Wuhan by January 22 (Appendix)."*
    - (Wu et al. 2020a) - *"In our baseline scenario, we estimated that $R_0$ was 2·68 (95% CrI 2·47–2·86) with an epidemic doubling time of **6.4** days (95% CrI 5·8–7·1; figure 2). We estimated that 75815 individuals (95% CrI 37 304–130 330) individuals had been infected in Greater Wuhan as of Jan 25, 2020."*
    - (Chinazzi et al. 2020) - *"Here we report the results for $T_g$ = 7.5 days (20). The obtained posterior distribution provides an average $R_0$ = 2.57 [90% confidence interval (CI): 2.37 to 2.78] and a doubling time of Td = **4.2** days (90% CI: 3.8 to 4.7 days)."*
    - (Q. Zhao, Chen, and Small 2020) - *"Assuming the travel rate to the selected countries and regions is constant over the study period, we found that the epidemic was doubling in size every **2.9** days (95% credible interval [CrI], 2 days--4.1 days). Using previously reported serial interval for 2019-nCoV, the estimated basic reproduction number is 5.7 (95% CrI, 3.4--9.2)"*.

■ [(Sanche et al. 2020)](#) - *"By using 2 distinct approaches, we estimated the growth rate of the early outbreak in Wuhan to be 0.21–0.30 per day (a doubling time of **2.3−3.3** days)"*

Caveats:
- If the fraction of cases that get detected changes over time, for example due to increased testing, it can bias estimates. Various alternative data sources such as the number of deaths can be used to compensate for varying ascertainment rates.
- Studies make different assumptions about when exponential growth began (estimates range from the end of November to 31th of December 2019), which leads to wide variation in estimates of the growth rate and doubling time.
- Our analysis (available [here](#)) suggests that most of the wide variation in the exponential growth rate, doubling times and reproduction number across studies of the Wuhan outbreak stems from the choice of the start date. Large exponential growth rates (and short doubling times) are mainly the result of assumptions regarding late initial time of exponential growth.
- It is most difficult to estimate the magnitude of the outbreak in its early phases. The number of reported cases is not necessarily reliable, as case count data can be confounded by low surveillance intensity causing a deflation in the true number of cases.
- In some cases inference was performed using international flight data and infected persons reported outside of China. Because of the low numbers of people traveling abroad compared with the total population size in Wuhan, this approach leads to substantial uncertainties. Inferences based on a low number of observations are prone to measurement error when data are incomplete or model assumptions are not fully justified. Furthermore, it was shown (Niehus et al. 2020) that in most countries there was underdetection of infected travellers returning from Wuhan (by a factor of close to 3 on average, when compared to the surveillance in Singapore).
- [(Q Zhao et al. 2020)](#) describes in detail how various elusive statistical biases can arise in models that infer the epidemic progression, usually resulting in underestimations.
- The growth rate is connected to the reproduction number via the generation interval. Similar estimates for the growth rate may result in different reproduction numbers depending on different assumptions regarding the generation interval and its distribution. See for example (J. Wallinga and Lipsitch 2007).

## Basic reproduction number

Definition
The basic reproduction number of an infection, $R_0$, is the expected number of secondary infections generated by an average infectious case in an entirely susceptible population.
This quantity determines the potential for an infectious agent to start an outbreak, the extent of transmission in the absence of control measures, and the ability of control measures to reduce spread. In contrast to $R_0$, the effective reproduction number, $R_t$ (also denoted R(t) and $R_e$), measures the number of secondary cases generated by an infectious case once an epidemic is underway (see the designated section for further details).

Best estimate: **2-6** (corresponding to doubling time of 2.5-7 days)

Methods of measurement and estimation:
$R_0$ and $R_t$ values are derived from fitting epidemiological models to data about the pandemic. These data can include the number of reported cases, hospitalizations and deaths. Importantly, since $R_0$ values are the product of a fitting procedure, their values depend on both the model and the quality of the underlying data. One approach to estimating $R_0$ is based on fitting the cumulative number of cases during initial phases of the epidemic. To estimate $R_0$ this approach usually relies on the most basic connection between $R_0$ and the doubling time and growth rate (*r*). These two parameters are interrelated via the generation interval (see relevant section). For a characteristic value of the generation interval g (units of days), with an exponential growth rate r (units of days$^{-1}$) the connection is given by $R_0 = e^{rg}$. The epidemic grows by a factor of $R_0$ over each generation interval. In general, the generation interval is characterized by a distribution g(t), where t is the time since infection, and not a single value, and the connection to $R_0$ is expressed through the Lotka-Euler equation:
$\frac{1}{R_0} = \int_{t=0}^{\infty} e^{-rt} g(t) dt = M(-r)$. Here, M is the moment generating function of g(t), which has a familiar form for commonly used distributions. For a generation interval with a normal distribution $R_0 = e^{r\mu - \frac{1}{2}(r\sigma)^2}$ where, $\mu, \sigma$ are the mean and standard deviation of the generation interval (J. Wallinga and Lipsitch 2007). This is also a good approximation for generation intervals following a gamma distribution.
Other approaches use a compartmental model to estimate $R_0$. Here $R_0$ is derived from the rates connecting the different compartments. For example, in an SIR (Susceptible-Infected-Removed) model, $R_0 = \frac{\beta}{\gamma}$ where $\beta$ is the effective contact rate, and $\gamma$ is the removal rate, equal to the inverse of the period of infectiousness. For general compartment models (such as SEIR) $R_0$ is given by the highest eigenvalue of the generation matrix (also known as the spectral radius).
Various sources estimate $R_0$ using the growth in the first stages of the outbreak and inference of the relevant rates from comparison to the number of infected. Comparison to various proxy data (exported cases, number of deaths) are used to validate and improve fits.

Estimates:
- Outbreak in Wuhan and mainland China:
    - Estimations on the basis of the number of cases exported from Wuhan to cities outside in mainland China and other countries:

- (Wu et al. 2020a) - SEIR model - *"In our baseline scenario, we estimated that R0 was **2.68** (95% CrI 2.47–2.86) with an epidemic doubling time of 6·4 days (95% CrI 5.8–7.1)"*. (assuming serial interval of 8.4, based on SARS)
- (Read et al., 2020) - SEIR model - *"We calculated the basic reproductive number, $R_0$, of the infection to be **3.11** (95%CI, 2.39–4.13)"*
- (Imai et al., n.d.) - Stochastic simulations of outbreak trajectories - *"We judge that the most likely estimate corresponds to the smallest level of zoonotic exposure explored here (40 cases), namely $R_0$=**2.6** (Table 1 and Figure 1). Uncertainty caused by the intrinsically random nature of epidemics and the uncertainty in the level of zoonotic exposure gives a range of 1.5-3.5...Assuming a shorter generation time (mean of 6.8 days rather than 8.4 days) reduces our central estimate of $R_0$ to **2.1** (uncertainty range: 1.3-2.7)"*
- (Zhanwei Du et al. 2020) - SEIR - *"Basic reproductive number, $R_0$: **1.90** (95% CrI 1.47–2.59)"* (given in the appendix, they got epidemic doubling time of 7.3 days)
- (Sanche et al. 2020) - integration of high-resolution domestic travel data and early infection data reported in provinces other than Hubei - *"To include a wider range of serial interval (i.e., 6–9 days) (Figure 5, panel A; Appendix 2 Figure 6), given the uncertainties in these estimations, we estimated that the median of estimated R is **5.7** (95% CI of 3.8–8.9) (Figure 5, panel B). The estimated R can be lower if the serial interval is shorter"* (They estimated that the exponential growth rate of the outbreak is probably 0.21–0.3/day. Used high serial interval)
- (Chinazzi et al. 2020) - *"Here we report the results for Tg = 7.5 days (20). The obtained posterior distribution provides an average $R_0$ = **2.57** [90% confidence interval (CI): 2.37 to 2.78] and a doubling time of Td = 4.2 days (90% CI: 3.8 to 4.7 days)."*

- Estimations on the basis of the reported number of cases:
    - (Shen et al. 2020) - SEIJR model (SEIR + isolated compartment) *"The basic reproduction number ($R_0$) of 2019-nCov, an indication of the initial transmissibility of the virus, was estimated to be **4.71** (4.50-4.92) when the epidemic started on 12th December 2019"* (used low rate of removal - inverse of time to isolation = 6 days)
    - (T. Liu et al. 2020) - exponential growth rate - *"The $R_0$ values of NCP were **4.5** (95%CI: 4.4-4.6) and 4.4 (95%CI: 4.3-4.6) nationwide and in Wuhan, respectively"* (assuming serial interval with mean of 7.5 and SD of 3.4)
    - (Riou and Althaus 2020) - Stochastic simulations of outbreak trajectories - *"the early human-to-human transmission of 2019-nCoV was characterised by values of $R_0$ around **2.2** (median value, with 90% high density interval: 1.4–3.8)"*
    - (Li et al. 2020) - exponential growth rate - *"Using the serial interval distribution above, we estimated that $R_0$ was **2.2** (95% CI, 1.4 to 3.9)"*.(the shorter generation time is closer to estimates of SARS-Cov2)
    - (Tang et al. 2020) - modified SEIR (including isolation) - *"Likelihood-based estimation of $R_c$ during the outbreak in Wuhan gives a mean value of **6.39** with mean and variance of generation time of 6 and 2 days on the basis of a revised data series (dataRev1). The reproduction number based on likelihood-based estimation ranges from 1.66 to 10"* (They assume exponential growth started at 1 Jan, resulting in a very high growth rate. Doubling time was about 2 days. High number of parameters in their model)

- [(S. Zhao, Lin, et al. 2020)](#) - exponential growth *"In Table 1, we estimated that the $R_0$ ranges from **2.24** (95%CI: 1.96-2.55) to **5.71** (95%CI: 4.24-7.54) associated with an 8-fold to 0-fold increase in the reporting rate"*
            - [(S. Zhao, Musa, et al. 2020)](#) - exponential growth with correction for underreporting of cases  -  *"After accounting for the effect of under-reporting, the $R_0$ was estimated at **2.56** (95% CI: 2.49–2.63), see Figure 1b, which is consistent with many existing online preprints with range from 2 to 4 [5,20–22]."*
    - Beginning of Spreading in other countries:
        - South Korea and Italy - [(Zhuang et al. 2020)](#) - *"We modelled the transmission process in the Republic of Korea and Italy with a stochastic model, and estimated the basic reproduction number $R_0$ as **2.6** (95% CI: 2.3–2.9) or **3.2** (95% CI: 2.9–3.5) in the Republic of Korea, under the assumption that the exponential growth starting on 31 January or 5 February 2020, and **2.6** (95% CI: 2.3–2.9) or **3.3** (95% CI: 3.0–3.6) in Italy, under the assumption that the exponential growth starting on 5 February or 10 February 2020, respectively"* - assumed short serial interval distribution - gamma with mean 4.5 and standard deviation of 3.1 days - might be good estimate for Korea due to their aggressive policy, but not necessarily for the beginning of the epidemic in Italy.
        - Iran - [(Muniz-Rodriguez et al. 2020b)](#) - *"To determine the transmission potential of severe acute respiratory syndrome coronavirus 2 in Iran in 2020, we estimated the reproduction number as **4.4** (95% CI 3.9–4.9) by using a generalized growth model and **3.5** (95% CI 1.3–8.1) by using epidemic doubling time."*  - assumed short serial interval distribution: gamma with mean 4.5 and standard deviation of 3.1 days, despite no aggressive interventions.

Comparison to other pathogens:
- SARS - [(Lipsitch et al. 2003)](#) - *"an approximate estimate of $R_0$ ranging from **2.2** to **3.6** for serial intervals of 8 to 12 days"*
- MERS - [(Majumder et al. 2014)](#) - *"IDEA model fits suggested a higher $R_0$ in Jeddah **(3.5-6.7)** than in Riyadh **(2.0-2.8)**; control parameters suggested more rapid reduction in transmission in the former city than the latter"*.
- Influenza - [(Biggerstaff et al. 2014)](#) - *"The median R value for 1918 was **1.80** (interquartile range [IQR]: 1.47–2.27). Six studies reported seven 1957 pandemic R values. The median R value for 1957 was **1.65** (IQR: 1.53–1.70). Four studies reported seven 1968 pandemic R values. The median R value for 1968 was **1.80** (IQR: 1.56–1.85). Fifty-seven studies reported 78 2009 pandemic R values. The median R value for 2009 was **1.46** (IQR: 1.30–1.70) and was similar across the two waves of illness: 1.46 for the first wave and 1.48 for the second wave. Twenty-four studies reported 47 seasonal epidemic R values. The median R value for seasonal influenza was **1.28** (IQR:1.19–1.37)"*.

Caveats:
- It is most difficult to estimate the magnitude of the outbreak in its early phases. The number of reported cases is not necessarily a good source, as case count data can be confounded by low surveillance intensity causing a deflation in the true number of cases. Another source for inconsistency is incidents originated by contact with  Wuhan seafood market rather than contact with other infected individuals.

- In some cases inference was performed by using international flight data and infected persons reported outside of China. Because of the low numbers of persons traveling abroad compared with the total population size in Wuhan, this approach leads to substantial uncertainties. Inferences based on a low number of observations are prone to measurement error when data are incomplete or model assumptions are not fully justified. Furthermore, it was shown (Niehus et al. 2020) that in most countries there was underdetection of infected travellers returning from Wuhan (by factor of close to 3 in average, when compared to the surveillance in Singapore).
- [(Q Zhao et al. 2020)](#) describes in detail how various elusive statistical biases can arise in models that infer the epidemic progression, usually resulting in underestimations.
- The reproduction number is connected to the serial interval and growth rate. The early sources didn't have reliable estimates for those, and thus are affected by the quality of their assumptions.
- Our analysis suggests that most of the wide variation in the exponential growth rate and reproduction number across studies of the Wuhan outbreak stems from the choice of the start day. Large reproduction numbers are mainly due to assumptions regarding late initial time of exponential growth.

**Effective reproduction number**

Definition
The effective reproduction number, $R_t$ (also denoted by R(t), R and $R_e$) is the number of secondary cases generated by an infectious case at a given moment (t) once an epidemic is underway. Unlike the basic reproductive number, $R_t$ is time and situation specific, used to characterize pathogen transmissibility during an epidemic, and enabling assessment of the effectiveness of interventions.

We should also emphasize the distinction between the 'physical' versus 'operational' meaning of $R_t$: the 'physical' meaning of the number of infections an infectious person makes on average (as defined above) does not depend on the generation interval. However, the operational definition which is inferred from models fitted to the number of cases or derived from logarithmic slope of infection curves depends on serial interval.

Best estimate: situation specific (depends on location and interventions)

Methods of measurement and estimation:
Like the basic reproduction number, $R_t$ values may be derived from fitting epidemiological compartmental models to data about the pandemic. These data can include the number of reported cases, hospitalizations and deaths. The two basic types of models are compartment-based models, such as SIR and SEIR, and renewal-equation models (see details below).
Estimation of effective $R_t$ via compartment models is similar to the estimation of $R_0$, and it makes use of the assumptions built into the model, such as: the different compartments and the distribution of time in each compartment. In order to estimate the effects of different interventions, such epidemiological models are also used to predict the $R_t$ based on assumptions and observed epidemiological parameters. Other approaches for estimation of the effective reproduction number use only the serial/generation interval via the renewal function, like those of (Cori et al. 2013; Jacco Wallinga and Teunis 2004). Renewal-equation estimates themselves can be divided into two basic methods, which differ by the inference method, as well as the use of right censored data:
- The first method (Jacco Wallinga and Teunis 2004) makes use of both past and future daily cases respective to the time step at which $R_t$ is inferred. Thus, it estimates the average number of secondary cases that a case infected at time step t eventually infected, estimated in retrospect. This method is also referred to as the "case reproduction number".
- The second method (Cori et al. 2013; Thompson et al. 2019) only uses past daily cases. Hence, the method estimates the average number of secondary cases that each infected individual would infect if the conditions remained as they were at time t. This method is also referred to as "the instantaneous reproduction number".

Estimates:
- Intensive interventions : isolation of patients with symptoms, contact tracing, physical distancing, including restriction (or advising against) of large social gatherings in order to mitigate the risk of community transmission:
  - Singapore (Tariq et al. 2020) - characterized daily incidence of local cases using the generalized logistic growth model (GLM) after adjusting for imported cases: *"The effective reproduction number peaked with a mean value **~1.0** between February 6-12, 2020 and declined thereafter. As of March 5, 2020, our most recent estimate of $R_t$ is at **0.9** (95% CI: 0.7,1.0)"*

- South Korea [(Shim et al. 2020)](#) - generalized growth model: *"... the intrinsic growth rate (r) was estimated at 0.6 (95% CI: 0.6, 0.7) and the scaling of growth parameter (p) was estimated at 0.8 (95% CI: 0.7, 0.8), indicating sub-exponential growth dynamics of COVID-19 in Korea (Figure 2, Table 3). The mean reproduction number $R_t$ was estimated at **1.5** (95% CI: 1.4, 1.6) as of Feb 26 2020."*
    - Hong Kong [(Cowling et al. 2020)](#) - Using the method by (Thompson et al. 2019), improved version of the approach of (Cori et al. 2013) *"Although unlinked COVID-19 cases have been detected in increasing numbers since early March, transmissibility ($R_t$) remains around the critical threshold of **1** (figure 2)"*
    - Sweden - no lock down [(Folkhalsomyndigheten 2020)](#) - using SEIR method - *"Based on reported cases during the period 17 February to 10 April, we calculated the value of the reproduction number at a certain time, $R_e(t)$. In all three scenarios, $R_e(t)$ takes a value **just above or just below one** on 10 April. See Table 3 for reproductions numbers with 95% confidence interval."*
- Interventions in Wuhan - before and after lockdown
    - [(Lipsitch M, Keya J, Cobey S.E. 2020)](#) following [(Pan et al. 2020)](#) - using (Jacco Wallinga and Teunis 2004) method during the period of local lockdown (period 3) $R_t$ dropped below 1 and reached 0.3 by the end of March - *"taking into account the evident decline of the epidemic curve that began during Period 3, finds that $R_t$ dropped below 1 during Period 3, specifically on 28-29 Jan. (Fig. 1)"*
    - [(C. Wang et al. 2020)](#) - extended SEIR model to account for population movement, unascertained cases, and quarantine by hospitalization - *"The effective reproductive number dropped from **3.86** (95% credible interval 3.74 to 3.97) before interventions to **0.32** (0.28 to 0.37) post interventions"*
- Interventions in other countries - before and after lockdown
    - France [(Salje et al. 2020)](#) - deterministic compartmental model stratified by age - from figure 4, $R_0$=3.3 and $R_{lockdown}$=0.5
    - 11 European countries [(Flaxman et al., 2020)](#) - semi-mechanistic Bayesian hierarchical model attempts to infer the impact of the various interventions (assuming each has the same relative effect on R, independent of the country). Based on reported death cases - *"We estimate large changes in $R_t$ in response to the combined non-pharmaceutical interventions. Our results, which are driven largely by countries with advanced epidemics and larger numbers of deaths (e.g. Italy, Spain), suggest that these interventions have together had a substantial impact on transmission, as measured by changes in the estimated reproduction number $R_t$. Across all countries we find current estimates of $R_t$ to range from a posterior mean of **0.97** [0.14-2.14] for Norway to a posterior mean of **2.64** [1.40-4.18] for Sweden, with an average of **1.43** across the 11 country posterior means, a 64% reduction compared to the pre-intervention values"*

Caveats:
- Various interventions such as isolation of symptomatic patients and contact tracing affect both the average number of secondary cases generated by an infected individual as well the serial interval itself.Usually the models assume a constant serial interval and consider only possible changes in $R_t$. Thus the models infer the 'operational' $R_t$ that correspond to a constant serial interval.
- Similar to the basic reproduction number, the effective reproduction number also depends on the quality of the data regarding the number of cases taken as is from formal reports, or estimated based on proxies such as traveling data.
- Other caveats relevant to the basic reproduction number are also relevant to the time specific statistics:
    - It is most difficult to estimate the magnitude of the outbreak in its early phases. The number of reported cases is not necessarily a good source, as case count data can be confounded by reservoir spillover events, stochasticity in the initial phase of the outbreak, and low surveillance intensity.
    - inference was performed by using international flight data and infected persons reported outside of China. Because of the low numbers of persons traveling abroad compared with the total population size in Wuhan, this approach leads to substantial uncertainties . Inferences based on a low number of observations are prone to measurement error when data are incomplete or model assumptions are not fully justified. Furthermore, it was shown (Niehus et al. 2020) that in most countries there was underdetection of infected travellers returning from Wuhan.

Comparison to other pathogens:
- SARS -
    - [(Wallinga and Teunis 2004)](#) - *"During the early phase of the SARS epidemic… Each case produced approximately three secondary infections (table 1). A value of R slightly higher than **3**… Around mid-March, control measures were implemented in all regions, and during this period the effective reproduction numbers decreased sharply… After the first World Health Organization global alert, each case produced approximately **0.7** secondary infections"*
    - [(Cori et al. 2013)](#) - *"For the SARS outbreak in Hong Kong in 2003, we find 2 successive peaks in R. The first occurred in the middle of week 3 with a median estimate of **12.2** (95% CI: 10.0–14.7), and the second occurred at the end of week 6 with a median estimate of **2.6** (95% CI: 2.4–2.9). R then fell below 1 by the end of week 7."*
- MERS - [(Thompson et al. 2019)](#) - fig 5: $R_t$ fluctuated between 10/14 to 1/16 between maximal value of less than 2.5, minimal value of about 0.5.
- Influenza - [(Thompson et al. 2019)](#) *"The median reproduction number estimate for the first seven days of the outbreak (April 18th–April 24th 2009) was **3.3** – with 95% credible interval (95% CrI) given by (2.1,5.6) – and the mean estimate for this period was 3.4"*.
- Ebola -
    - [(Althaus 2015)](#) - using SEIR model - *"I fitted an EVD transmission model to previously published data of this outbreak and estimated the basic reproduction number $R_0$ = 5.2 (95% CI [4.0–6.7]). The model suggests that the net reproduction number $R_t$ fell **below unity** 28 days (95% CI [25–34] days) after the onset of symptoms in the index case"*. See figure 2 for details.
    - [(Thompson et al. 2019)](#) - fig 4: estimates for $R_t$ in Liberia depend on the assumption of serial interval distribution's shape.

Online tools to estimate the effective reproduction number in various locations
- https://mrc-ide.github.io/covid19-short-term-forecasts/#individual-model-outputs
- https://covid19-projections.com/infections-tracker/
- http://rt.live/
- https://mrc-ide.github.io/covid19estimates/#/ (https://github.com/ImperialCollegeLondon/covid19model)
- http://trackingr-env.eba-9muars8y.us-east-2.elasticbeanstalk.com/
- https://covidactnow.org/

**Dispersion of the secondary infection distribution**

Definition
The "offspring distribution" of an epidemic is the distribution of the number of secondary cases due to each primary infection. The mean of this distribution is $R_t$ by definition, though $R_t$ values derived from fits might differ from this true $R_t$ value due to incorrect assumptions or issues associated with the fitting process. The offspring distribution is generally considered to adopt a negative binomial distribution which is characterized by a mean ($R_t$) and a dispersion parameter k, such that the variance is given by $R_t(1+R_t/k)$. The negative binomial distribution is generally used instead of Poisson to allow for unequal mean and variance. Smaller values of k indicate greater heterogeneity. In general k<<1 is associated with a high number of 'superspreaders' - individuals that infect much more than the average $R_0$. Values of k>1 are associated with a low number of 'superspreaders', as the number of secondary cases of each primary infection is relatively even.

The formula for the probability density p of the number of secondary infections n, is given by:
$p(n) = \frac{\Gamma(n+k)}{n!\,\Gamma(k)}\left(\frac{k}{k+R_t}\right)^k\left(\frac{R_t}{k+R_t}\right)^n$ where $R_t$ is the mean number of secondary cases, k is the dispersion of the secondary infection distribution, and $\Gamma$ denote the gamma function. The limits of this distribution are the Poisson distribution, when k→∞ and the geometric distribution when k=1.

Best estimate:
No interventions: **0.1-0.2**
Interventions (physical distancing, contact tracing, lockdowns) : **0.4-1**

Methods of measurement and estimation:
Estimation is made based on a chain transmission model, in which each infected person transmits the infection to a number of subjects sampled randomly from a distribution with mean $R_t$ and dispersion k. Usually, the model is compared to data regarding the size of different clusters (for different countries, for example) with an initial number of spreaders (for example, known imported cases).

Other methods use empirical data regarding the number of secondary cases in a defined cluster with good documentation. They fit the transmission data to a negative binomial distribution to extract an estimate for the dispersion parameter. A version of this method, proposed by (Lloyd-Smith et al. 2005) makes use of only the average number of secondary cases and the proportion of cases with no secondary transmission by numerically solving an equation that connects the probability of no-transmission with the dispersion.

Estimates:

- Early outbreak:
    - Early spreading in Wuhan - (Riou and Althaus 2020) - *"The observed data at this point are compatible with a large range of values for the dispersion parameter k (median: 0.54, 90% high density interval: 0.014–6.95). However, our simulations suggest that very low values of k are less likely."*
    - Early spreading in several countries - (Endo et al. 2020) - *"The result of the joint estimation suggested the likely bounds for $R_0$ and k (95% CrIs: $R_0$ 1.4–12; k **0.04−0.2**)"*.
- Containment effort - physical distancing and contact tracing:

- - China, Shenzhen - (Bi et al. 2020) - contact tracing - *"The mean number of secondary cases caused by each index case (i.e., the observed reproductive number, R), was 0.4 (95% CI 0.3,0.5). The distribution of personal reproductive numbers was highly overdispersed, with 80% of infections being caused by 8.9% (95% CI 3.5,10.8) of cases (negative binomial dispersion parameter **0.58**, 95% CI 0.35, 1.18)."*
  - Hong Kong under strong containment - (Adam et al. 2020)- *"From the empirical offspring distribution and fitted negative binomial distribution shown in Figure 2B, we estimated an observed reproductive number (R) of 0.58 (95% CI: 0.45 – 0.71) and dispersion parameter (k) of **0.45** (95% CI: 0.31 – 0.76). Given a superspreading of 6-8 secondary cases, we directly observed 2-4 SSEs where the source was known. Likelihood analysis based on final cluster sizes increased the estimate of R to 0.75 (95% CI: 0.6 – 0.96) and decreased estimates of k to **0.37** (95% CI: 0.16 – 1.16). Based on these estimates we determined a coefficient of variation of 2.5 and inferred that approximately 20% of SARS-CoV-2 infections are responsible for 80% of all transmission events in Hong Kong"*
  - Singapore under strong containment - (Tariq et al. 2020) - *"The dispersion parameter is estimated at **0.4** (95% CI: 0.1, Inf.) indicating higher transmission heterogeneity"*
- Tianjin, China before and after containment efforts - (Y Zhang et al. 2020) - *"we estimated the reproductive number R and the dispersion parameter k (lower value indicating higher heterogeneity) to be 0.67 (95% CI: 0.54–0.84) and **0.25** (95% CI: 0.13–0.88), respectively"*. They also give results for before (k=0.14) and after government control measures (k=0.77)
- China outside of Hubei Province under travel lockdown - (D. He et al. 2020), (Xu et al. 2020) - *"we obtained raw data of number of secondary cases from a study by Xu et al [4], where k can be readily calculated as **0.8** (95%CI 0.63, 0.98)."*
- Diamond princess cruise ship - (S. Zhao, Cao, et al. 2020) - *"We estimate the dispersion term (k) to be **44** (95%CI: 6–88), which is significantly larger than 1 and consistent with [7]. The simulation results with or without dispersion term (k) were largely consistent, which suggests absence of superspreading events (note that low k implies high chance of superspreading events)"*. Because this is only anecdotal evidence from a non-representative setting, we do not include it in our best estimate

Caveats:
- Some of the analysis relies on the overall size of clusters and their initial number of spreaders. However, in an emerging disease it is hard to assess the total size of the cluster as it depends on the fraction of the population being tested (and the testing policy). In addition, the initial number of (imported) cases may be influenced by misdetection of a few cases.
- It is not always clear from case reports which cases are epidemiologically linked. Hence, the extent and precision of the epidemiological investigation could also affect the assessment of cluster sizes and the proportion of cases with no secondary infections.
- Many of the interventions meant to control the spreading, like isolation of symptomatic patients and contact tracing, affect the dispersion, as they aim to prevent events of large spreading (as well as the mean = $R_0$). Thus, it is important to divide the data according to the level of intervention

Comparison to other pathogens:
- (Lloyd-Smith et al. 2005) - analyzed previous directly transmitted infections (Fig. 1b, c and Supplementary Tables 1, 2):

SARS - **0.16** (90%CI 0.11-0.64)
Measles (US 1997-1999) - **0.23** (0.16-0.39)
Smallpox (Europe, Benin, Pakistan) - **0.32-0.72**
- MERS - [(Kucharski and Althaus 2015)](#) - *"Our estimate for MERS-CoV is similar, with k = **0.26** (90% CI: 0.11– 0.87, 95% CI: 0.09–1.24)"*.
- Influenza (1918 Pandemic) - [(Fraser et al. 2011)](#) - **0.94** (0.59-1.72) (taken from table 1)

**Fraction of asymptomatic infections**

Definition: The fraction of infected people not showing noticeable symptoms throughout their infection. Note, this is different from pre-symptomatic patients, who carry the virus but will only show symptoms later (once their incubation period has passed).

Best estimate: **20-70%, a population-wide average, with strong age dependence such that estimates depend on demographics.**

Methods of measurement and estimation:
There are three main methods for estimating the fraction of asymptomatic infections. The first method randomly samples the population to identify the subset of infected individuals, and then follows-up on those cases to see if they develop symptoms.

The second relies on contact tracing of index cases in specific clusters, identification of the infected contacts of the index cases and monitoring of their symptoms. This type of measurement is not a fully random sample and might exhibit some biases in the sampled population. We refer to this type of measurement as "semi-random".

The third method relies on epidemiological compartment models. In these models, the fraction of asymptomatic cases can be represented by a unique compartment separate from the symptomatic cases. The fraction of asymptomatic cases is one parameter in this model, which could be fitted, i.e. inferred, along with the other parameters of the model, to time course data of the epidemic.

Estimates:
- Random sample
  - Iceland - 1221 cases ([Gudbjartsson et al. 2020](#)) - "*Notably, **43%** of the participants who tested positive reported having no symptoms, although symptoms almost certainly developed later in some of them*"
  - Vo, Italy - 81 cases - ([Lavezzo et al. 2020](#)) - "*Notably, **43.2%** (95% CI 32.2-54.7%) of the confirmed SARS CoV-2 infections detected across the two surveys were asymptomatic*"
  - Tibetan Autonomous Prefecture - 83 cases - [(Song et al. 2020)](#) - "*18 (18/83, **21.7**%) individuals were identified as asymptomatic carriers, with a predominant distribution of males (61.0%) (Table 1). The median age was 31 years and one third of the asymptomatic individuals were students, aged under 20 years*".
  - Spain - about 3000 seropositive cases - [(Pollán et al. 2020)](#) - "*Around a third of seropositive participants were asymptomatic, ranging from **21.9%** (19.1–24.9) to **35.8%** (33.1–38.5).*"
- Contact tracing (semi-random)
  - Korea, call center - 97 cases - [(Park et al. 2020)](#) - "*The role of asymptomatic COVID-19 case-patients in spreading the disease is of great concern. Among 97 confirmed COVID-19 case-patients in this study, 4 **(4.1%)** remained asymptomatic during the 14-days of monitoring*" (72%) were women; mean age was 38 years (range 20–80 years)
  - Shenzhen, China - 391 cases - [(Bi et al. 2020)](#) "*Further, in this group, **20%** were asymptomatic at the time of first clinical assessment and nearly 30% did not have fever*".
  - Brunei - 71 cases - [(Chaw et al. 2020)](#) - "*A substantial proportion of cases were presymptomatic (31.0%, n=22) or asymptomatic (**12.7%**, n = 9)*"

- - Lombardy, Italy - 2351 close contacts of index cases - (Poletti et al. 2020) - *"Of the 2,351 confirmed SARS-CoV-2 infections, 849 (36.1%) were symptomatic."*, i.e. **63.9**% asymptomatic, *"Data stratified by sex, age and testing procedure are displayed in Table 1."*
  - Taiwan - 100 cases - (Cheng et al. 2020) - *"As of March 18, 2020, there were 100 patients with laboratory-confirmed COVID-19 in Taiwan, including 10 clusters of patients and 9 asymptomatic patients."* i.e. the fraction of asymptomatic cases is **9%**
  - Guangzhou, China - 129 cases - (Luo et al. 2020) - *"During quarantine period, 129 cases (2.6%) were diagnosed, with 8 asymptomatic (**6.2%**), 49 mild (38.0%), and 5 (3.9%) severe to critical cases"*

- Epidemiological models
  - Wuhan - (Davies et al. 2020) - *"The age-dependent clinical proportion was markedly lower in younger age groups in all regions (Fig 2b), with 20% of infections in children under 10 resulting in clinical cases, rising to over 70% in adults over 70 in the consensus age distribution estimated across all regions."* Thus, 80% asymptomatic cases in children under 10 and 30% in adults over 70. A simple average over the averages of the age groups in Table 3 yields an average of **57%**. Non clinical cases include asymptomatic as well as mild cases, not reported or documented.

Caveats:
- Reporting of mild symptoms is subjective and so the dividing line between pre-symptomatic and asymptomatic cases is not well-defined.
- While in some cases, a large collection of samples were taken, it is very challenging to take a truly random sample of a large population. Many studies report the asymptomatic fraction for a highly idiosyncratic study population that certainly differs greatly from the general population, for example the passengers of the Diamond Princess cruise ship..
- To estimate the true fraction of asymptomatic cases, researchers need to follow-up on cases which did not show symptoms at the time of testing to verify that these are truly asymptomatic and not presymptomatic cases.
- Even though there is evidence that asymptomatic cases have viral loads similar to those of symptomatic cases (Zou et al. 2020), there are, on the other hand, also reports of a connection between the severity of symptoms and viral load (Liu et al. 2020). Thus there are contradictory reports on the relationship between symptom severity and the viral load, so it is unclear if asymptomatic cases would be harder to detect in random surveys. If asymptomatic cases indeed go undetected, this will cause an underestimation of their true fraction.
- When performing random population testing, the false positive and false negative rates of the test need to be accounted for in order to accurately project the associated uncertainty of the estimates based on the tests. This may lead to significant over-estimation of asymptomatic cases in screening tests, where the disease prevalence is very low.
- A recent review (Oran and Topol 2020) states in its abstract, based on data largely overlapping with the sources provided above, that a range of 40%-45% of infections are asymptomatic. Such a narrow confidence range is hard to support given the large variation in the fraction of asymptomatic infections between studies, as well as other possible biases and caveats described above. In the discussion it is stated less conclusively: *"On the basis of the 3 cohorts with representative samples—Iceland and Indiana, with data gathered through random selection of participants, and Vo', with data for nearly all residents—the asymptomatic infection rate may* *be as high as 40% to 45%. A conservative estimate would be 30% or higher to account for the presymptomatic admixture that has thus far not been adequately quantified"*
- It is important to make the distinction between the fraction of asymptomatic cases and the fraction of undiagnosed, undetected or unreported cases. In seroprevalence studies, the actual fraction of infected cases is much higher (many times more than 10-fold) than the number of reported cases (see the IFR section for examples of such seroprevalence studies). This does not mean that all these extra cases are all asymptomatic. Even when an infected person shows symptoms, there is still a large probability that their infection will not be tested or reported, for different reasons (symptoms might not be significant enough to make the infected person seek medical help, limitations on the testing capacity, etc.).

Lower quality data

- Small sample sizes:
  - Japanese citizens evacuated from Wuhan, China - 13 cases - ([Nishiura et al. 2020](#))- *"Using a binomial distribution, the asymptomatic ratio is thus estimated at **30.8%** (95% confidence interval (CI): 7.7%, 53.8%)"* (sample size was very small).
  - Korea - 28 cases [Ki et al. 2020](#) - *"Of the 28 infected patients diagnosed in Korea, 3 were asymptomatic"*
  - Hubei - 23 cases [Long et al. 2020](#) - *"A total of 16 out of 164 cases were confirmed by RT-PCR during Jan 31 to Feb 9, 2020, with 3 cases reporting no symptoms. The other 148 cases were no symptoms and negative in RT-PCR tests. On March 1, 2020, serum samples were collected from these 164 cases for antibody tests. The 16 RT-PCR confirmed cases were positive in IgG or/and IgM. Strikingly, 7 of the 148 cases who were excluded previously by negative nucleic acid results also showed positive results in IgG or/and IgM, indicating that 4.3% (7/162) of close contacts were missed by nucleic acid test. In addition, about 6.1% (10/164) of this cohort were asymptomatic infection."* - There were 10 asymptomatic cases among a total of 23 cases identified, implying **43%** are asymptomatic.
  - Vietnam - 30 cases [Chau et al. 2020](#) - *"Of these, 30 participated in the study: 13(**43%**) never had symptoms and 17(57%) were symptomatic"*
  - Liaocheng, Chine - 24 cases [Tian et al. 2020](#) - Table 1*"Asymptomatic (%) 7/24 (**29.17**)"*
- Non representative sample:
  - USA, Nursing home in Washington - [Arons et al. 2020](#) - **6%** (3/48) *"Of the 48 residents who tested positive from the surveys, 17 (35%) reported typical symptoms, 4 (8%) reported only atypical symptoms, and 27 (56%) reported no new symptoms or changes in chronic symptoms at the time of testing… In the 7 days after their positive test, 24 of the 27 asymptomatic residents (89%) had onset of symptoms and were recategorized as presymptomatic"* - mean age of 76, 98% with coexisting conditions.
  - Diamond Princess cruise ship - ([Mizumoto et al. 2020](#)) - delay adjusted - *"The estimated asymptomatic proportion was **17.9%** (95% credible interval (CrI): 15.5–20.2%)"* - older population.
  - Children in China ([Dong et al](#)) - *"For the severity of patients (including both confirmed and suspected cases), 94 (**4.4 %**), 1091 (50.9 %) and 831 (38.8 %) patients were diagnosed as asymptomatic, mild or moderate cases, respectively, totally accounted for 94.1 % of all cases."* - Only children, not clear if suspected cases are indeed cases.

- - Repatriation flights to Greece 40 cases - ([Lytras et al. 2020](#)) - Table 1. 35 asymptomatic cases out of 40 positive cases or **87.5%.** *"The passengers' median age was 27 years (interquartile range 22–40 years)"*.
  - Pregnant women admitted for delivery [(Sutton et al. 2020)](#) - *"Thus, 29 of the 33 patients who were positive for SARS-CoV-2 at admission (**87.9%**) had no symptoms of Covid-19 at presentation."*
  - Homeless shelter in Boston, USA [(Baggett et al. 2020)](#) - *"A total of 147 participants (36.0%) had PCR test results positive for SARS-CoV-2. Men constituted 84.4% of individuals with PCR-positive results and 64.4% of individuals with PCR-negative results. Among individuals with PCR test results positive for SARS-CoV-2, cough (7.5%), shortness of breath (1.4%), and fever (0.7%) were all uncommon, and **87.8%** were asymptomatic."*

**Household secondary attack rate**

Definition: The fraction of the infectee's household members that they infect on average.

Best estimate: **5-30%**

Methods of measurement and estimation: The secondary attack rate is usually estimated based on contact tracing data. By identifying the primary infected individual in the household, other household members are tested and secondary infections are identified. The secondary attack rate is calculated as the fraction of the household members of an infected individual who end up infected with the virus.

Estimates:
- [(Bi et al. 2020)](#) - *"The household secondary attack rate was **11·2%** (95% CI 9·1–13·8), and children were as likely to be infected as adults."*
- [(Jing et al. 2020)](#) - *"We estimated the household SAR to be **13.8%** (95% CI: 11.1-17.0%) if household contacts are defined as all close relatives and **19.3%** (95% CI: 15.5-23.9%) if household contacts only include those at the same residential address as the cases"*
- [(Wang et al. 2020)](#) - *"The rate of secondary transmission among household contacts of patients with SARS-CoV-2 infection was **30%***"
- [(Park et al. 2020)](#) - *"The household secondary attack rate among symptomatic case-patients was **16.2%** (95% CI 11.6%– 22.0%)"*
- [(WHO 2020)](#) - *"Household transmission studies are currently underway, but preliminary studies ongoing in Guangdong estimate the secondary attack rate in households ranges from **3-10%**"*
- [(Korean CDC 2020)](#) - *"There were 119 household contacts, of which 9 individuals developed COVID-19 resulting in a secondary attack rate of **7.56%** (95% CI 3.7–14.26)"*
- [(Burke et al. 2020)](#) - *"and a symptomatic secondary attack rate of **10.5%** (95% CI = 2.9%−31.4%) among household members."*
- [(Cheng et al. 2020)](#) - *"The attack rate was higher among household (**4.6%** [95% CI, 2.3%-9.3%]) and nonhousehold (5.3% [95% CI, 2.1%-12.8%]) family contacts than that in health care or other settings"*
- [(Luo et al. 2020)](#) - *"Among different modes of contact, 53 household contacts were the most dangerous in catching with infection of 54 COVID-19, with an incidence of **10.2%**."*
- [(Chen et al. 2020)](#) - *"Living with the case **(13.26%)**, taking the same transportation (11.91%), dinner and entertainment (7.18%) are all high-risk factors for infection"*
- [(Li et al. 2020)](#) - *"Total 105 index patients and 392 household contacts were enrolled… The secondary attack rate of SARS-CoV-2 in household is **16.3%**."*
- [(Chaw et al. 2020)](#) - *"The overall household AR is **10.6%** (95% CI: 7·3, 15·1)"*

Caveats:
- The detection of cases following a contact tracing study may miss some of the infections if the timing of the study is in sync with the progression of infection of household members, and is subject to the detection error rates of the detection method.
- In some cases only clinical cases are identified, thus underestimating the true secondary attack rate.
- There could be cases in which there was either more than one infector in the household, or that some people in the household got infected from a source other than the household.
- The specific attack rate could change based on the awareness of the disease, hygiene and

physical distancing measures.

**Case fatality rate (CFR)**

Definition: The total number of fatalities divided by the total number of confirmed cases

Best estimate: **1-15%**
(uncorrected for age structure, selection bias, ascertainment rate, time delays etc.)

Methods of measurement and estimation: Descriptive statistics based on summary of gathered data. Possible biases not taken into account in this estimate are specified in the caveats section.

Estimates:
- [Global country statistics from Worldometer analyses by our open script](#) (accessed on 31/05/2020) - We use data from all countries with more than 50 death cases and calculate the uncorrected raw Case Fatality Rate for each country. The range represents the lowest and highest rates observed.

Caveats:
Caveats associated with estimating the CFR can be divided into those affecting the numerator, i.e. the total number of fatalities, and those affecting the denominator, i.e. the total number of cases:
- Caveats associated with estimating the number of fatalities:
  - Several key factors can affect the death toll. First, demographic parameters and practices associated with increased or decreased risk differ greatly across societies. For example, the capacity of the healthcare system, the average age of the population, and the prevalence of smoking. Indeed, the majority of people dying from SARS-CoV-2 have a preexisting condition such as a cardiovascular disease or smoking [(China CDC 2020)](#).
  - There is a potential for overestimating the CFR due to a tendency to identify more severe cases (selection bias) which will tend to overestimate the CFR.
  - There is usually a delay between the onset of symptoms and death, which can lead to an underestimate of the total number of deaths early in the progression of an epidemic.
  - Attribution of cases could sometimes be challenging, with some deaths unaccounted for due to lack of testing or due to registration under a different cause of death. One way to get around this issue is to quantify excess deaths relative to the average number of deaths over the last several years.
  - Different studies apply different types of corrections for a subset or all of the above mentioned caveats. For example see (Wu et al. 2020).
- Caveats associated with estimating the number of cases:
  - The total number of cases detected in each country is highly dependent on the testing capacity in the specific country. In places where death statistics are reliable, this bias would cause countries with limited testing capacity to have a higher CFR than countries with better testing capacity.

**Infection fatality rate (IFR)**

Definition: The total number of fatalities divided by the total number of infected people

Best estimate: **0.3-1%**

Methods of measurement and estimation:
Estimating the total number of infected people, and thus the IFR, requires testing a random sample of the population.

The main methods used for estimating the IFR are:
- Testing of semi-random samples using RT-qPCR
- Testing of semi-random samples via serology (antibody) testing
- Model-based estimates

Because people who were infected and recovered do not show as positive with viral RNA quantification methods, the ideal way to estimate the number of infected people is testing for the presence of antibodies (serosurvey) against SARS-CoV-2 in random samples. As serosurveys were not available in the early stages of the pandemic, researchers resorted to semi-random samples tested with qPCR, or to modelling-based estimates. As of writing (May 2020), several serosurveys are being conducted and results with IFR estimates are starting to be released.

Estimates:
- Semi-random samples using RT-PCR:
    - Repatriation flights from China - (Verity et al. 2020) - *"We obtain an overall IFR estimate for China of **0.66**% (0.39%,1.33%)"* and (Ferguson et al. 2020) - *"The IFR estimates from Verity et al.12 have been adjusted to account for a non-uniform attack rate giving an overall IFR of **0.9%** (95% credible interval 0.4%-1.4%)."*
    - Diamond Princess cruise ship - (Russel et al. 2020) - *"We estimated that the all-age cIFR on the Diamond Princess was **1.3%** (95% confidence interval (CI): 0.38–3.6) "*
- Model-based estimates:
    - Estimates for early epidemic in China (Mizumoto et al 2020) - *"crude infection fatality ratio (IFR) and time−delay adjusted IFR is estimated to be 0.04% (95% CrI: 0.03%−0.06%) and **0.12%** (95%CrI: 0.08−0.17%)"*
    - France (Salje et al. 2020) - *"We find 2.6% of infected individuals are hospitalized and **0.53%** die, ranging from 0.001% in those <20y to 8.3% in those >80y. "* based on data from the Diamond Princess cruise ship incorporated into a statistical model.
    - Global (Grewelle et al. 2020) - *"Applying this asymptotic estimator to cumulative COVID-19 data from 139 countries reveals a global IFR of **1.04%** (CI: 0.77%,1.38%)."*
- Serosurveys:
    - Germany - (Streeck et al. 2020) - *"Die Letalität (case fatality rate) bezogen auf die Gesamtzahl der Infizierten in der Gemeinde Gangelt beträgt mit den vorläufigen Daten aus dieser Studie ca. **0,37 %.**"*
    - Denmark  - (Erikstrup et al. 2020) - *"Using available data on fatalities and population numbers a combined IFR in patients younger than 70 is estimated at **82 per 100,000** (CI: 59-154) infections"* To translate the IFR for people under 70 to the total IFR, please consider this twitter tread
    - Santa-Clara, CA, USA - (Bendavid et al. 2020) - Due to methodical problems with this study the results are not indicative and thus not presented.

- Unpublished serosurveys
  - New York State - [link](#) (doesn't give an IFR estimate yet).
  - Los Angeles County - [link](#) (doesn't give an IFR estimate yet).
  - Spain - [link](#) (doesn't give an IFR estimate yet).
  - Indiana - [link](#) - "*IUPUI scientists estimate the infection-fatality rate for the novel coronavirus in Indiana to be **0.58 percent***"

Caveats:
Caveats associated with estimating the IFR can be divided into those affecting the numerator, i.e. the total number of fatalities, and those affecting the denominator, i.e. the total number of infected people.
- Caveats associated with estimating the number of fatalities:
  - Demographic parameters and practices associated with increased or decreased risk differ greatly across societies. For example, the capacity of the healthcare system, the average age of the population, and the prevalence of smoking. Indeed, the majority of people dying from SARS-CoV-2 have a preexisting condition such as cardiovascular disease or smoking [(China CDC 2020)](#).
  - There is usually a delay between the onset of symptoms and death, which can lead to an underestimate of the total number of deaths early in the progression of an epidemic.
  - Attribution of cases could sometimes be challenging, with some deaths unaccounted for due to lack of testing or due to registration under a different cause of death.
- Caveats associated with estimating the total number of infected people:
  - To get a good estimate of the prevalence infected people using a random and representative sample of the population must be taken. In many cases, achieving such random sampling is very challenging, and in most cases it is hard to determine how representative the sample is. For example, people who have experienced the hallmark symptoms of SARS-CoV-2 may be more inclined to enroll in an open study.
  - When the prevalence of a disease is low, false positive results may skew the results significantly. Many of the tests done rely on methods with non-negligible false positive and false negative rates, and thus the reported estimates may have significant uncertainty associated with them, if the false positive rate of the test was not accounted for appropriately.
  - The detection of infected individuals using PCR based tests is sensitive to the time at which the test was performed relative to the progression of the disease in infected individuals. In the first days of infection or close to recovery, the viral load of infected individuals can be under the detection limit of PCR tests and thus infections might be missed.
- Caveats specific to serosurveys:
  - The usage of a small number of defined SARS-CoV-2 proteins in antibody tests, e.g. spike and/or nucleoprotein, can result in underestimates.
  - The detection of infected individuals is sensitive to the time at which the test was performed relative to the progression of the disease. It requires about 2 weeks on average for antibodies to reach detectable levels.

**Probability of hospitalization**

Definition: The fraction of infected individuals that need to be hospitalized due to their conditions.

Best estimate: **2-4% of infected** (hospitalization rate out of confirmed cases could be much higher, e.g. 20%, as most mild cases can remain untested)

Methods of measurement and estimation:
Descriptive statistics based on summary of gathered hospitalization data. Often divided by age groups, gender and comorbidities. This value must be scaled to the total number of infected people, similar to the rescaling of CFR to infer IFR. Estimation of the number of infected people requires testing a random sample of the population. Because people who were infected and recovered do not show as positive with viral RNA quantification methods, the ideal way to estimate the number of infected people is testing for the presence of antibodies (serosurvey) against SARS-CoV-2 in random samples. As these were not available in the early stages of the pandmeic, researchers resorted to semi-random samples tested with qPCR, or to modelling-based estimates. Current estimates rely on semi-random samples, used to correct hospitalization data.

Estimates:
- The fraction of infected individuals that need to be hospitalized (based on semi-random samples):
    - Repatriation flights from Wuhan, China (adjusted for under-ascertainment rates by comparison to other locations in China):
        - (Verity et al. 2020) - "*The demography-adjusted and under-ascertainment-adjusted proportion of infected individuals requiring hospitalisation ranges from **1.1%** in the 20–29 years age group up to **18.4%** in those 80 years and older*" (Table 3 for age distribution. Assuming a uniform attack rate. Hospitalization of only severe patients).
        - (Ferguson et al., 2020) - "*The age-stratified proportion of infections that require hospitalisation and the infection fatality ratio (IFR) were obtained from an analysis of a subset of cases from China (Verity et al.). These estimates were corrected for non-uniform attack rates by age and when applied to the GB population result in an IFR of 0.9% with **4.4%** of infections hospitalised (Table 1)*" (assuming hospitalization of only severe patients))
    - Vo, Italy - (Lavezzo et al. 2020) - "*Of the 81 SARS-CoV-2 positive patients across the two surveys, 14 required hospitalization (**17.2%**). Their age distribution was as follows: 1 (7.1%) in the 41-50 age group, 2 (14.3%) in the 51-60 age group, 4 (28.6%) in the 61-70 age group, 5 (35.7%) in the 71-80 age group and 2 (14.3%) in the 81-90 age group*"
    - Synthesis of total infected cases from Diamond Princess with ICU and hospitalization data from France - (Salje et al. 2020) - "*We find that **2.6%** of infected individuals are hospitalized (95% CrI: 1.4-4.4), ranging from 0.09% (95% CrI: 0.05-0.2) in females under <20y to 31.4% (95% CrI: 16.7-52.6) in males >80y.*" (see table for distribution for sex and age).

- The fraction of confirmed cases that need to be hospitalized (based on surveys)

- China - (Wu and McGoogan 2020) - *"Most cases were classified as mild (81%; ie, non pneumonia and mild pneumonia). However, **14%** were severe (ie, dyspnea, respiratory frequency 30/min, blood oxygen saturation 93%, partial pressure of arterial oxygen to fraction of inspired oxygen ratio <300, and/or lung infiltrates >50% within 24 to 48 hours), and **5%** were critical (ie, respiratory failure, septic shock, and/or multiple organ dysfunction or failure)"*
- USA
    - (Bajema et al. 2020) - *"All 210 persons who were tested were symptomatic… Forty-two (**20%**) patients were hospitalized, and four (2%) were admitted to an intensive care unit."*
    - Louisiana - (Price-Haywood et al. 2020) - *"A total of **39.7%** of Covid-19–positive patients (1382 patients) were hospitalized, 76.9% of whom were black"* they conclude: *"Black race was not associated with higher in-hospital mortality than white race, after adjustment for differences in sociodemographic and clinical characteristics on admission"*.
    - Among Children in the USA - (CDC COVID-19 Response Team 2020) - *"Information on hospitalization status was available for 745 (29%) cases in children aged <18 years and 35,061 (31%) cases in adults aged 18–64 years. Among children with COVID-19, 147 (estimated range = **5.7%−20%**) were reported to be hospitalized, with 15 (0.58%−2.0%) admitted to an ICU (Figure 2)"*. see figure for age distribution

Caveats:
Caveats associated with estimating the probability of hospitalization can be divided into those affecting the numerator i.e. the total number of individuals need to be hospitalized and those affecting the denominator, i.e. the total number of infected people.

- Caveats associated with estimating the number of fatalities:
    - Patients in need of hospitalization can be in different conditions. Health systems in places which suffer from a high number of infections would tend to hospitalize only patients with severe conditions, but in other situations patients with milder conditions would also be hospitalized.
    - Hostpitalization probability is highly dependent on age and sex, thus the overall hospitalization in a country is affected by its population age structure, as well as the different fraction of infected individuals (also known as "attack rate") among different ages.
    - Underlying conditions and comorbidity are factors affecting the severity of the disease. Their prevalence is correlated with age, but differs between countries. Thus, it may alter the probabilities, and make it difficult to estimate it correctly for other countries.

- Caveats associated with estimating the total number of infected people in studies regarding early stages of the epidemic:
    - Non random sampling, common at early stages of the epidemic.
    - Reliance on a small number of individuals to assess the fraction of infected individuals is problematic, especially when it is further divided into several age groups. Specifically, the sample size of known cohorts for young ages (<20y) is very small.

Comparison to other pathogens:
- SARS - [(Rainer et al. 2004)](#) - very few percent do not need hospitalization "Of 910 patients who were managed without hospitalization, 6 patients had serologic evidence of SARS. Five of the 6 patients had normal chest radiographs, and 4 patients had symptoms such as myalgia, chills, coughing, and feeling feverish".

**Probability of ICU admission**

Definition: The fraction of hospitalized individuals that are admitted to the Intensive Care Unit (ICU). The probability of ICU admission is also often regarded as the fraction of hospitalized patients in critical condition (i.e., respiratory failure, septic shock, and/or multiple organ dysfunction or failure).

Best estimate: **10-40%** (large variation between health systems).

Methods of measurement and estimation:
Descriptive statistics based on summary of gathered hospitalization data. Often divided by age groups, gender and comorbidities.

Estimates:
- China, Wuhan
  - (D. Wang et al. 2020) - **26%** of hospitalized were admitted to ICU *"Compared with patients who did not receive ICU care (n=102), patients who required ICU care (n=36) were significantly older (median age, 66 years [IQR, 57-78] vs 51 years (IQR, 37-62); P<.001) and were more likely to have underlying comorbidities"*
  - (Huang et al. 2020) - *"Of the 41 patients, 13 (**32%**) were admitted to the ICU because they required high-flow nasal cannula or higher-level oxygen support measures to correct hypoxaemia"*
  - (Yang et al. 2020) - **27%** of hospitalized were critical (55 of 201 included**)** - see figure 1.
- China - (Wu and McGoogan 2020) - *"Most cases were classified as mild (81%; ie, non pneumonia and mild pneumonia). However, 14% were severe (ie, dyspnea, respiratory frequency 30/min, blood oxygen saturation 93%, partial pressure of arterial oxygen to fraction of inspired oxygen ratio <300, and/or lung infiltrates >50% within 24 to 48 hours), and 5% were critical (ie, respiratory failure, septic shock, and/or multiple organ dysfunction or failure)"* The estimate of **25%** is based on dividing the number of critical cases by the number of severe cases (similar to critical out of hospitalized, as mostly only severe cases are hospitalized in a crowded health system)
- Lombardy Region, Italy - (Grasselli, Pesenti, and Cecconi 2020) - *"The proportion of ICU admissions represents 12% of the total positive cases, and **16%** of all hospitalized patients. This rate is higher than what was reported from China, where only 5% of patients who tested positive for COVID-19 required ICU admission"*.
- France, analysis of probability by age, based on data combined with incidents from Diamond Princess cruise ship - (Salje et al. 2020) - *"Once hospitalized, **18.2%** (95% CrI: 18.0%-18.6%) patients enter ICU after a mean delay of 1.5 days (Figure S1). There is an increasing probability of entering ICU with age - however, this drops for those >70y (Figure 2B, Table S2).."* (see table for distribution for sex and age).
- New York - (Richardson et al. 2020) - from Table 5, **22%** of hospitalized patients (1281/5700) were admitted to the ICU.
- China, Shanghai - (Chen et al. 2020) - *"A total of 22 (**8.8%**) patients were admitted in ICU 8.5 ±4.0 days after onset of symptoms"*.
- USA (end of January) - (Bajema et al. 2020) - **10%** of hospitalized were admitted to ICU. *"Forty-two (20%) patients were hospitalized, and four (2%) were admitted to an intensive care unit."*
- Louisiana, USA - (Price-Haywood et al. 2020) - *"**More than one third** of admitted patients (474 patients) received care in an intensive care unit, among whom 80.2% were black patients."* from table 2: 29.5% of White Non-Hispanic patients and 35.7% of Black Non-Hispanic patients

- Georgia, USA - (Gold et al. 2020) - *"Intensive care unit (ICU) admission occurred among 119 (**39.0%**) patients and increased significantly with age group: among patients aged ≥65 years, 53.8% were admitted to an ICU (p<0.001). Overall, 92 (30.2%) patients received IMV, representing 77.3% of those admitted to an ICU."* Similar fractions among Whilat and Black patients.
- Specific for children:
  - China - (Dong et al. 2020) - *"The proportions of severe and critical cases were 10.6%, 7.3%, 4.2%, 4.1%, and 3.0% for the age groups ,1, 1 to 5, 6 to 10, 11 to 15, and >16 years, respectively...with much fewer being severe and critical cases (5.8%) than in adult patients (18.5%)"* Based on Table 2, we estimate a probability of **10%** for ICU admissions from the ratio of critical cases to cases in severe or critical condition
  - USA - (CDC COVID-19 Response Team 2020) about **10%** - *"Information on hospitalization status was available for 745 (29%) cases in children aged <18 years and 35,061 (31%) cases in adults aged 18–64 years. Among children with COVID-19, 147 (estimated range = 5.7%–20%) were reported to be hospitalized, with 15 (0.58%–2.0%) admitted to an ICU (Figure 2)"*.
    (see figure for age distribution)

Caveats:
- Patients in need of hospitalization can be in different physical conditions. Health systems in places which suffer from a high number of infections would tend to hospitalize only patients with severe conditions and thus have higher probability of ICU admission.
- Both hostpitalization and ICU admission probabilities are highly dependent on age and sex, thus the overall hospitalization of a country is affected by its population age structure, as well as the different attack rate among different ages.
- Underlying conditions and comorbidity are factors affecting the severity of the disease. Their prevalence is correlated with age, but differs between countries. Thus, it may alter the probabilities, and make it difficult to estimate it correctly for other countries.

Comparison to other pathogens:
- SARS - (Joynt and Yap 2004) - *"Approximately **20%** of patients with severe acute respiratory syndrome (SARS) develop respiratory failure that requires admission to an intensive care unit (ICU)"*.
- MERS - (Matsuyama et al. 2016) - *"The risks of ICU admission and mechanical ventilation ranged from 44% to 100% and from 25% to 100%, with pooled estimates at **78%** and 73%, respectively"*

**Probability of death in ICU**

Definition: The fraction of infected patients that were admitted to the Intensive Care Unit (ICU) who didn't survive.

Best estimate: **20%-80%**

Methods of measurement and estimation:
Descriptive statistics based on summary of gathered hospitalization data. Often divided by age groups, gender, ethnicity and comorbidities.

Estimates:
- In health systems crowded by the outbreak:
  - China, Wuhan
    - (D. Wang et al. 2020) - **17%** of patients admitted to ICU have died *"Of the 36 patients admitted to the ICU, 11 were still in the ICU, 9 had been discharged to home, 10 had been transferred to the general wards, and 6 had died".*
    - (Huang et al. 2020) - *"13 (32%) patients were admitted to an ICU and six (**15%**) died".*
    - (Yang et al. 2020) - *"32 (**61.5%**) of critically ill patients had died at 28 days. Of all included patients, 37 (71%) required mechanical ventilation and 35 (67%) had ARDS."*
  - New York, USA - (Richardson et al. 2020) - According to Table 5, **78%** (291/373) of the ICU patients who had outcomes (been discharged or died) have died. Due to the overcrowding of the health system, a similar number of patients have died without being admitted to the ICU - *"Among the 2634 patients who were discharged or had died at the study end point, during hospitalization, 373 (14.2%) were treated in the ICU, 320 (12.2%) received invasive mechanical ventilation, 81 (3.2%) were treated with kidney replacement therapy, and 553 (**21%**) died (Table 5)."*
  - Lombardy Region, Italy - (Grasselli et al. 2020) - **61%** of ICU patients with outcomes have died (405/661). 26% of admitted (including those stille in ICU). This fraction may be overestimated due to censoring effects. *"Among the 1581 patients with ICU disposition data available as of March 25, 2020, 920 patients (58% [95% CI, 56%-61%]) were still in the ICU, 256 (16% [95% CI, 14%-18%]) were discharged from the ICU, and 405 (26% [95% CI, 23%-28%]) had died in the ICU. Older patients (n = 786; age 64 years) had higher mortality than younger patients (n = 795; age 63 years) (36% vs 15%; difference, 21% [95% CI, 17%-26%]; P < .001)"*
  - Britain - (ICNARC 2020c) - from Table 10, **43%** (3481/8060) have died in critical care (out of those with outcomes). See table for separation by age, ethnicity , gender and more.
- In less crowded health systems:
  - China, Shanghai - (Chen et al. 2020) - **22%** of ICU patients with outcomes have died (2/9)."*A total of 22(8.8%) patients were admitted in ICU... To the time of submission, a total of 215 (86.3%) patients were discharged after 16(12–20) days hospitalization, 13 patients were still in ICU and 19 patients in stable condition. A total of 2 patients died (0.8%)."*.
  - Georgia, USA - (Gold et al. 2020) - *"As of April 24, 2020, 24 (7.9%) patients remained hospitalized, including 14 (58.3%) in the ICU and nine (37.5%) on IMV. Overall, the estimated percentage of deaths among patients who received ICU care ranged from*

- ***37.0%**, assuming all remaining ICU patients survived, to **48.7%**, assuming all remaining ICU patients died".*
- Probability of death of <u>hospitalized</u> patients:
  - France, analysis of probability by age, based on data combined with incidents from Diamond Princess - [(Salje et al. 2020)](#) - *"Overall, **18.1%** (95% CrI: 17.8–18.4) of <u>hospitalized</u> individuals go on to die"*. Very close to the fraction of admitted to ICU (18.2%).
  - Louisiana, USA - [(Price-Haywood et al. 2020)](#) - from table 2, **25%** of <u>hospitalized</u> patients with outcomes have died (326/1320). Among White patients about 31% die, while only 22% among Black patients.

<u>Caveats</u>:
- Patients in need of hospitalization can be in different conditions. Health systems in places which suffer from a high number of infections would tend to hospitalize only patients with severe conditions and thus have higher probability of ICU admission. Moreover, overcrowding of ICU may result in critical patients not being admitted to ICU at all.
- Censoring of ICU outcomes (i.e. outcomes being available for only part of the patients during the spread) may bias the death probability to higher values as the duration of stay in ICU is longer for patients who recover. This is also shown in previous reports from ICNARC ([ICNARC 2020a](#), [ICNARC 2020b](#)).
- Hostpitalization, ICU admission and death probabilities are highly dependent on age and sex, thus the overall hospitalization in a country is affected by its population age structure, as well as the different attack rate among different ages groups.
- Underlying conditions and comorbidity are factors affecting the severity of the disease. Their prevalence is correlated with age, but differs across countries.

<u>Comparison to other pathogens</u>:
- SARS - [(Joynt and Yap 2004)](#) - *"Long-term mortality for patients admitted to the ICU ranges from **30% to 50%**."*.
- MERS - [(Matsuyama et al. 2016)](#) - From fig 4, among four studies that consider only ICU cases, the death probability ranged from **58% to 75%**.

**RT-qPCR False Positive Rate**

Definition: The fraction of uninfected people who test positive for SARS-CoV-2 via RT-qPCR. The false positive rate is often reported as "specificity" which is the fraction of negatives test results that are true negatives (i.e. uninfected). The false positive rate (in %) is 100% - specificity.

Best estimate: 0.2-3% corresponding to a specificity of 97-99.8%.

Methods of measurement and estimation: Testing of samples from patients known not to have SARS-CoV-2 based on various modes of testing, patients who could not have had SARS-CoV-2 due to sample collection prior to late 2019, and samples from animals known not to contract SARS-CoV-2.

Estimates:
- RT-qPCR testing of uninfected patients and/or animals
  - (Jiang et al. 2020) - *"The sensitivity and specificity of RT-PCR for pharyngeal were 78.2% and 98.8%. Positive predictive value and negative predictive value were 91.9% and 96.2. and Youden's index was 0.770, which indicated overall performance."* A specificity of 98.8% corresponds to a false positive rate of **1.2%**.
  - (Wernike et al. 2020) "If we assume a best-case scenario for specificity based on these results for the A-3 or B / E-Sarbeco setting, the diagnostic specificity was calculated as 0.9756 (40/41; table 1)." A specificity of 0.9756 corresponds to a false positive rate of **2.44%**.
- Comparison to established RT-qPCR testing for other RNA viruses
  - (Merckx et al. 2017) Table 2 gives a pooled specificity of 99.2% (95% confidence interval of 98.6-99.7%) for influenza A and 99.4% (98.9-99.8) for influenza B. A specificity of 99.2% and 99.4% corresponds to a false positive rate of **0.8%** and **0.6%**.
  - (Cohen and Kessel 2020) "In all, 336 of the 10,538 negative samples (3.2%) were reported as positive. We considered two data sets comprising all 43 EQAs (full data set), and the 37 EQAs that analyzed at least 100 negative samples (subset). FPRs in each EQA ranged from 0-16.7% for the full data set, and 0- 8.1% for the subset... The median and interquartile range were lower for the full data set (median=**2.3%**, interquartile range=0.8-4.0%) than for the subset (median=2.5%, interquartile range=1.2-4.0%)."

Caveats:
- RT-qPCR is used to quantify the amount of viral genomic RNA in patient samples. It works by monitoring the progress of a reverse-transcriptase (RT) PCR via the fluorescence of a DNA binding dye (or an oligonucleotide probe). Since DNA accumulates over the course of the reaction and dye fluorescence is increased by DNA binding, fluorescence reports on reaction progress. It is common to report "cycle threshold" or "quantification cycle" values ($C_t$ or $C_q$ values) for RT-qPCR. $C_q$ values report the cycle number at which dye fluorescence crossed a predefined threshold. The earlier the threshold is crossed, the lower the $C_q$ value and the more RNA was in the sample. A positive test is one with a $C_q$ value beneath a prespecified threshold (e.g. Ct < 40), though this threshold will vary from between specific tests (e.g. which primers were used).
- The false positive rate or specificity of RT-qPCR testing is only rarely reported in studies investigating RT-qPCR tests of SARS-CoV-2.

- Error rates can depend strongly on the primer sets used, as documented in (Wernike et al. 2020) and (Barra et al. 2020), the latter of which is quoted here: *"We observed consistent false positive results for N and N2 (60 out of 60 samples for both). N3 assay generated false positive signal or inconclusive results in 13 out of 60 tested samples."*
- False positives are likely to be dominated by errors in sample handling, as false positives are uncommon in RT-qPCR in general.

**RT-qPCR False Negative Rate**

Definition: The fraction of people infected with SARS-CoV-2 who test negative by a single RT-qPCR test. The false negative rate is often reported as "sensitivity" which is the fraction of negative test results that are true positives (i.e. were previously infected). The false negative rate (in %) is therefore 100% - sensitivity.

Best estimate: 5-30**%** depending on choice of kit and primers

Methods of measurement and estimation: RT-PCR testing of confirmed SARS-CoV-2 patients in the days immediately following the onset of symptoms. Patients are typically identified via an initial positive RT-qPCR test and a second diagnostic method, e.g. a CT scan of the lungs.

Estimates:
- RT-qPCR testing of confirmed SARS-CoV-2 patients
    - (J. Zhao Jr. et al. 2020) Figure 2, Table 2 and results - *"As the results shown in Figure 2 and Table 2, in the early phase of illness within 7-day since onset, the RNA test had the highest sensitivity of 66.7%, whereas the antibody assays only presented a positive rate of 38.3%."* Sensitivity of 66.7% corresponds to a false negative rate of **33.3%**.
    - (Jiang et al. 2020) - *"Combination of RT-PCR and CT has the higher sensitivity (91.9%,79/86) than RT-PCR alone (78.2%, 68/87) or CT alone (66.7%, 54 of 81) or combination of two RT-PCR tests (86.2%,75/87)."* Sensitivity of 78.2% corresponds to a false negative rate of **22.8%**.
    - (Arnaout et al. 2020) - *"For example, an assay with LoD of 1,000 copies/mL, such as that of the CDC assay (8) or Genmark ePlex EUA (9), is expected to detect 77%, or 3 in 4, of infected individuals, for a false-negative rate of **22%**. With an LoD of 6,250 copies/mL, the LabCorp COVID-19 RT-PCR EUA test has an estimated clinical sensitivity of 67% and a false-negative rate of **33%**, missing approximately 1 in 3 infected individuals"*
- Evaluation of multiple rounds of patient testing
    - (Li et al. 2020) "Among the 384 patients with initial negative results, the second test was performed. For these patients, the test results were positive in 48 cases (12.5%), dubiously positive in 27 patients (7.0%), negative in 280 patients (72.9%), and results were not available for 29 patients (7.6%) (Figure 1B and Table 2). Among the patients with initial non-positive results, seven patients were eventually confirmed with COVID-19 by three repeated swab PCR tests, four were confirmed by four repeated tests, and one was confirmed by five repeated tests." Simple calculation gives a nominal false negative rate of at least 12/384 ≈ **3%** and at most 48/384 ≈ **12%**.
- Meta-analysis of published test results
    - (Wikramaratna et al. 2020) - *"The probability of a positive test decreases with the number of days past symptom onset; for a nasal swab, the percentage chance of a positive test declines from 94.39% [86.88, 97.73] on day 0 to 67.15% [53.05, 78.85] by day 10."* This implies that the false negative rate increases from **5.61%** 0 days after symptom onset to **32.85%** 10 days after symptom onset
    - (Kucirka et al. 2020) "Over the 4 days of infection before the typical time of symptom onset (day 5), the probability of a false negative result in an infected person decreases from 100% (95% CI, 100% to 100%) on day 1 to 67% (CI, 27% to 94%) on day 4, although there is considerable uncertainty in these numbers. On the day of symptom onset, the median false-negative rate was **38%** (CI, 18% to 65%)."
- Comparison to established RT-qPCR testing for other RNA viruses

- (Merckx et al. 2017) Figure 3 gives a pooled sensitivity of 92% (95% confidence interval of 85–96%) for influenza A and 95% (87-99%) for influenza B. Sensitivity of 92% and 95% corresponds to a false negative rate of **8%** and **5%**.

Caveats:
- RT-qPCR is used to quantify the amount of viral genomic RNA in patient samples. It works by monitoring the progress of a reverse-transcriptase (RT) PCR via the fluorescence of a DNA binding dye (or an oligonucleotide probe). Since DNA accumulates over the course of the reaction and dye fluorescence is increased by DNA binding, fluorescence reports on reaction progress. It is common to report "cycle threshold" or "quantification cycle" values ($C_t$ or $C_q$ values) for RT-qPCR. $C_q$ values report the cycle number at which dye fluorescence crossed a predefined threshold. The earlier the threshold is crossed, the lower the $C_q$ value and the more RNA was in the sample. A positive test is one with a $C_q$ value beneath a prespecified threshold (e.g. Ct < 40), though this threshold will vary from between specific tests (e.g. which primers were used).
- There is an important distinction between empirical and analytical sensitivity. The analytical sensitivity is also known as the "limit of detection." The limit of detection is related to empirical sensitivity (equivalently, false negative rate) in that patient samples containing amounts of SARS-CoV-2 RNA beneath the limit of detection will nearly always produce false negative results. The limit of detection is measured by adding known quantities of synthetic SARS-CoV-2 RNA into an RT-qPCR reaction; the lowest measurable quantity is the limit of detection. These values are expressed in units of viral RNA copies per reaction and vary depending on which primers and kits are used (Vogels et al. 2020). However, analytical sensitivities are usually measured in the absence of patient sputum or saliva, which might affect PCR reactions. Moreover, for patient samples, the true quantity of RNA is always unknown and inter-individual variation in physiology and genetics may contribute to variation in viral titers in sputum or saliva. Empirical sensitivities and false negative rates reported above therefore represent RT-qPCR studies conducted on confirmed COVID-19 patients. However, since confirmed COVID-19 patients are nearly always symptomatic and often selected from hospital admissions, there may be a selection bias towards sicker individuals with higher viral loads, which would tend to yield under-estimates of the false negative rate. It is therefore very striking that reported false negative rates sometimes exceed 30%.
- The false negative rate of RT-qPCR testing appears to depend greatly on sample handling, storage and protocol details, for example the choice of primers and kits (Barra et al. 2020; Vogels et al. 2020), which might explain the wide range of estimates.
- Viral titers depend strongly on how much time has passed since infection. Generally, viral titers in patient samples increase over time until reaching a maximum at ≈5 days after infection, after which titers decrease again (Zhao et al. 2020; Wikramaratna et al. 2020; Larremore et al. 2020). When titers are beneath the limit of detection, one should not expect RT-qPCR to detect the infection. At the moment of infection, for example, patients sputum and saliva should test negative because there has been no time for the virus to replicate in their body. It is also important to note that this represents an inferred description of a "typical" patient and that there is certainly inter-individual variability in maximal viral load and sputum production that will affect individual likelihoods of diagnosis of RT-qPCR.
- Double testing of the same patient may substantially reduce the false negative rate and is recommended by the CDC and several references (Jiang et al. 2020). However, the quantitative effect of multiple samples of the same individuals usually lowers the false negative rate to a lesser degree than expected when tests are statistically independent. For example, (Jiang et al. 2020) report a 22% false negative rate for a single test and 14% false negative rate for a double test. If we assume statistical independence, then the false negative rate for two tests should be $0.22^2 \approx 5\%$. The empirical value of 14% represents an 8% reduction in the false negative rate, which is about ½ of the 17% reduction expected when tests are statistically independent. This lower-than-expected reduction could stem from the repeat tests not being completely independent, due, for example, to the timing of tests relative to the start of infection or the time the tests are taken during the time course of the infection of the infected individual.
- Similarly, confirmatory testing by CT scan (Jiang et al. 2020) or antibody tests (Zhao et al. 2020) reduce the false negative rate as well.

**Duration of PCR positivity**

Definition: The number of days post symptom onset for which >50% of infected people test positive by RT-qPCR. Notably, viral shedding and, hence, PCR-positivity, begins several days after infection.

Best estimate: ≈7-9 days

Methods of measurement and estimation: RT-PCR testing of confirmed SARS-CoV-2 patients in the days following the onset of symptoms. Patients are typically identified via an initial positive RT-qPCR test and potentially a second diagnostic method, e.g. a CT scan of the lungs.

Estimates:
- RT-qPCR testing of confirmed SARS-CoV-2 patients
    - (J. Zhao Jr. et al. 2020) Figure 2, Table 2 and results - *"As the results shown in Figure 2 and Table 2, in the early phase of illness within **7**-day since onset, the RNA test had the highest sensitivity of 66.7%, whereas the antibody assays only presented a positive rate of 38.3%."*
    - (Young et al. 2020) - *"Among the 18 hospitalized patients with PCR-confirmed SARS-CoV-2 infection (median age, 47 years; 9 [50%] women), clinical presentation was an upper respiratory tract infection in 12 (67%), and viral shedding from the nasopharynx was prolonged for **7** days or longer among 15 (83%)."*
    - (To et al. 2020) Figure 2 and text - *"Of the 21 patients who survived, seven (33%) had viral RNA detected for 20 days or longer after symptom onset."*
- Re-analysis of published PCR testing results
    - (Wikramaratna et al. 2020) Figure 3, which replots data from (Bi et al. 2020)

Caveats:
- The period of >50% PCR positivity may not begin immediately after onset of symptoms. For example, one reference report a ≈1 day lag from symptom onset to 50% PCR positivity (Zhao et al. 2020), but this reference had access to only one patient whose reported symptoms began on the same day as testing.
- Error rates associated with RT-qPCR testing will depend greatly on sample handling, storage and protocol details, which might explain the wide range of estimates.
- As with all PCR testing, error rates and limits of detection will depend on the choice of primers and kits (Barra et al. 2020; Vogels et al. 2020).

**Antibody Test False Positive Rate**

Definition: The fraction of uninfected people who test positive for SARS-CoV-2 antibodies. The false positive rate is often reported as "specificity" which is the fraction of negatives test results that are true negatives (i.e. uninfected). The false positive rate (in %) is 100% - specificity.

Best estimate: 0.1-15%, corresponding with a specificity of ≈85-99.8%.

Methods of measurement and estimation: testing of samples from patients known not to have SARS-CoV-2 because the samples were taken before the epidemic began (e.g. banked blood donations and patient sera).

Estimates:
- Testing of pre-epidemic samples
  - [(Whitman et al. 2020)](#) - *"Test specificity ranged from 84.3-100.0% in pre-COVID-19 specimens."* A specificity of 84.3%-100% corresponds to a false positive rate of **0%-15.7%**.
  - [(J. Zhao Jr. et al. 2020)](#) - *"The specificity of the assays for Ab, IgM and IgG was determined as 99.1% (211/213), 98.6% (210/213) and 99.0% (195/197) by testing of samples collected from healthy individuals before the outbreak of SARS-CoV-2"*. A specificity of 99.1%, 98.6% and 99% corresponds to a false positive rate of **0.9%**, **1.4%** and **1%**.
  - [(Bryan et al. 2020)](#) - *"We tested 1,020 serum specimens collected prior to SARS-CoV-2 circulation in the United States and found one false positive, indicating a specificity of 99.90%."* A specificity of 99.9% corresponds to a false positive rate of **0.1%**.
  - [(Jääskeläinen et al. 2020)](#) *"The specificity and sensitivity values of the commercial tests against MNT, respectively, were as follows: 95.1%/80.5% (Abbott Architect SARS-CoV-2 IgG), 94.9%/43.8% (Diasorin Liaison SARS-CoV-2 IgG), 68.3%/87.8% (Euroimmun SARS-CoV-2 IgA), 86.6%/70.7% (Euroimmun SARS-CoV-2 IgG), 74.4%/56.1% (Acro 2019-nCoV IgG), 69.5%/46.3% (Acro 2019-nCoV IgM), 97.5%/71.9% (Xiamen Biotime SARS-CoV-2 IgG), and 88.8%/81.3% (Xiamen Biotime SARSCoV-2 IgM)."* A specificity of 95.1%, 94.9%, 68.3% and 74.4%, 69.5%, 97.5% and 88.8% corresponds to a false positive rate of **4.9%, 5.1%** and **31.7%, 25.6%, 30.5%**, and **11.2%**. We consider kits with especially high values as outliers and not representative for the best estimate above.
- Comparison to established immunological testing for other RNA viruses
  - [(Merckx et al. 2017)](#) Table 2 gives a pooled specificity estimate of 99.4 (95% confidence interval 99.1-99.7%) for influenza A and 99.8% (99.7-99.9%) for influenza B. A specificity of 99.4%, and 99.8% corresponds to a false positive rate of **0.6%** and **0.2%**.

Caveats
- Serology tests assay the presence of antibodies in the bloodstream. As antibodies only develop ≈10 days after the onset of symptoms (Zhao et al. 2020; To et al. 2020) these tests are not useful for diagnosing patients with active disease. Rather, serology testing or "serosurveys" are used to measure the incidence of the infection in the population at large in order to determine the state of the epidemic. Because these tests have non-trivial false-positive and false-negative rates, great care should be taken in interpreting the results of serosurveys.

- Since antibodies take ≈10 days to develop. However, the timing of infection is often unknown. In practice, therefore, the timing of the test will be a major factor affecting test results.
- Error rates will depend both on which testing approach (or specific kit) is used as well as the sample collection, storage and sample handling procedures. We report a range that includes kit-to-kit variability as reported in (Whitman et al. 2020; Jääskeläinen et al. 2020) but this does not include between-lab differences in sampling, transport and preparation. Some kits have been shown to perform worse than the reported range, see (Jääskeläinen et al. 2020).

**Antibody Test False Negative Rate**

Definition: The fraction of previously infected people who test negative for SARS-CoV-2 antibodies. Since it takes time for antibodies to develop, we report a false negative rate estimate for tests conducted 10 or more days after symptoms. The false negative rate is often reported as "sensitivity" which is the fraction of negative test results that are true positives (i.e. were previously infected). The false negative rate (in %) is 100% - sensitivity.

Best estimate: 10-50%, corresponding to a selectivity of 50-90% 10+ days after symptom onset.

Methods of measurement and estimation: testing of samples from patients known to have SARS-CoV-2 by a combination of RT-qPCR and clinical markers (e.g. symptoms and CT scans). One study (Jääskeläinen et al. 2020) compares serology tests to assays testing the neutralization of virus by patient sera. In this particular case, we report numbers for patients irrespective of time post symptoms as the control (viral neutralization) directly tests for the presence of neutralizing antibodies.

Estimates:
- Testing of samples from patients with confirmed SARS-CoV-2 10+ days after onset of symptoms.
    - Table 2 of (Whitman et al. 2020) gives sensitivity as a function of days since patient-reported onset of symptoms. Note that the sample number is small in this study (< 30 for all bins).
    - (J. Zhao Jr. et al. 2020) Figure 2 and results section - *"In samples from patients during day 8-14 after onset, the sensitivities of Ab (89.6%), IgM (73.3%) and IgG (54.1%) were all higher than that of RNA test (54.0%)."* Sensitivity of 89.6% and 54.1% corresponds to a false negative rate of **10.4%** and **45.9%**.
    - (Bryan et al. 2020) - *"The sensitivity of the assay from the estimated day of symptom onset for the 125 patients included in our chart-review study was 53.1% (95% CI 39.4%-66.3%) at 7 days, 82.4% (51.0-76.4%) at 10 days, 96.9% (89.5-99.5%) at 14 days, and 100% (95.1%-100%) at day 17 using the manufacturer's recommended cutoff of 1.4."* Sensitivity of 82.4% and 96.9% corresponds to a false negative rate of **11.6%** and **3.1%**.
- Comparison to viral neutralization measurement
    - (Jääskeläinen et al. 2020) "The specificity and sensitivity values of the commercial tests against MNT, respectively, were as follows: 95.1%/80.5% (Abbott Architect SARS-CoV-2 IgG), 94.9%/43.8% (Diasorin Liaison SARS-CoV-2 IgG), 68.3%/87.8% (Euroimmun SARS-CoV-2 IgA), 86.6%/70.7% (Euroimmun SARS-CoV-2 IgG), 74.4%/56.1% (Acro 2019-nCoV IgG), 69.5%/46.3% (Acro 2019-nCoV IgM), 97.5%/71.9% (Xiamen Biotime SARS-CoV-2 IgG), and 88.8%/81.3% (Xiamen Biotime SARSCoV-2 IgM)." Sensitivity of 80.5% and 43.8%, 87.8%, 70.7%, 56.1%, 46.3%, and 71.9% corresponds to a false negative rate of **19.5%, 56.2%, 12.2%, 29.3%, 43.9%, 53.7%**, and **28.1%**. We consider kits with especially high values as outliers and not representative for the best estimate above.
- Comparison to established immunological testing for other RNA viruses
    - (Merckx et al. 2017) Table 2 gives a pooled sensitivity estimate of 54.4% (95% confidence interval 48.9-59.8%) for influenza A and 53.2% (41.7-64.4%) for influenza B.

Sensitivity of 54.4% and 53.2% corresponds to a false negative rate of **45.6%** and **46.8%**.

Caveats

- Serology tests assay the presence of antibodies in the bloodstream. As antibodies only develop ~10 days after the onset of symptoms (Zhao et al. 2020; To et al. 2020) these tests are not considered useful for diagnosing patients with active disease. Rather, serology testing or "serosurveys" are used to measure the incidence of the infection in the population at large in order to determine the state of the epidemic. Because these tests have non-trivial false-positive and false-negative rates, great care should be taken in interpreting the results of serosurveys.
- Since antibodies take ~10 days to develop. However, the timing of infection is often unknown. In practice, therefore, the timing of the test will be a major factor affecting test results.
- As above, error rates will depend both on which testing approach (or specific kit) is used as well as the sample collection, storage and sample handling procedures. We report a range that includes kit-to-kit variability as reported in (Whitman et al. 2020; Jääskeläinen et al. 2020) but this does not include between-lab differences in sampling, transport and preparation.
- The range we report is the empirical range from (Whitman et al. 2020; Jääskeläinen et al. 2020) and does not include statistical estimates of uncertainty. A statistical estimate of the 95% confidence interval on test sensitivity gives much wider ranges.

# References

(From definitions and caveates. Measurements & estimates have direct links to papers).

Data to Quantify Bias of Traveller-Derived COVID-19 Prevalence Estimates in Wuhan, China." *The Lancet Infectious Diseases*, April. https://doi.org/10.1016/S1473-3099(20)30229-2.

Oran, Daniel P., and Eric J. Topol. 2020. "Prevalence of Asymptomatic SARS-CoV-2 Infection: A Narrative Review." *Annals of Internal Medicine*, June. https://doi.org/10.7326/M20-3012.

Park, Sang Woo, Kaiyuan Sun, David Champredon, Michael Li, Benjamin M. Bolker, David J. D. Earn, Joshua S. Weitz, Bryan T. Grenfell, and Jonathan Dushoff. 2020. "Cohort-Based Approach to Understanding the Roles of Generation and Serial Intervals in Shaping Epidemiological Dynamics." *Infectious Diseases (except HIV/AIDS)*. medRxiv. https://doi.org/10.1101/2020.06.04.20122713.

Thompson, R. N., J. E. Stockwin, R. D. van Gaalen, J. A. Polonsky, Z. N. Kamvar, P. A. Demarsh, E. Dahlqwist, et al. 2019. "Improved Inference of Time-Varying Reproduction Numbers during Infectious Disease Outbreaks." *Epidemics* 29 (December): 100356.

To, Kelvin Kai-Wang, Owen Tak-Yin Tsang, Wai-Shing Leung, Anthony Raymond Tam, Tak-Chiu Wu, David Christopher Lung, Cyril Chik-Yan Yip, et al. 2020. "Temporal Profiles of Viral Load in Posterior Oropharyngeal Saliva Samples and Serum Antibody Responses during Infection by SARS-CoV-2: An Observational Cohort Study." *The Lancet Infectious Diseases*, March. https://doi.org/10.1016/S1473-3099(20)30196-1.

Vogels, Chantal B. F., Anderson F. Brito, Anne Louise Wyllie, Joseph R. Fauver, Isabel M. Ott, Chaney C. Kalinich, Mary E. Petrone, Marie-Louise Landry, Ellen F. Foxman, and Nathan D. Grubaugh. 2020. "Analytical Sensitivity and Efficiency Comparisons of SARS-COV-2 qRT-PCR Assays." *medRxiv*. https://www.medrxiv.org/content/medrxiv/early/2020/04/26/2020.03.30.20048108.full.pdf.

Wallinga, Jacco, and Peter Teunis. 2004. "Different Epidemic Curves for Severe Acute Respiratory Syndrome Reveal Similar Impacts of Control Measures." *American Journal of Epidemiology* 160 (6): 509–16.

Wallinga, J., and M. Lipsitch. 2007. "How Generation Intervals Shape the Relationship between Growth Rates and Reproductive Numbers." *Proceedings. Biological Sciences / The Royal Society* 274 (1609): 599–604.

Wernike, Kerstin, Markus Keller, Franz J. Conraths, Thomas C. Mettenleiter, Martin H. Groschup, and Martin Beer. 2020. "Pitfalls in SARS-CoV-2 PCR Diagnostics." *Transboundary and Emerging Diseases*, June. https://doi.org/10.1111/tbed.13684.

Whitman, Jeffrey D., Joseph Hiatt, Cody T. Mowery, Brian R. Shy, Ruby Yu, Tori N. Yamamoto, Ujjwal Rathore, et al. 2020. "Test Performance Evaluation of SARS-CoV-2 Serological Assays." *Infectious Diseases (except HIV/AIDS)*. medRxiv. https://doi.org/10.1101/2020.04.25.20074856.

Wikramaratna, Paul, Robert S. Paton, Mahan Ghafari, and Jose Lourenco. 2020. "Estimating False-Negative Detection Rate of SARS-CoV-2 by RT-PCR." *medRxiv*. https://www.medrxiv.org/content/10.1101/2020.04.05.20053355v2.abstract.

Wu, Joseph T., Kathy Leung, Mary Bushman, Nishant Kishore, Rene Niehus, Pablo M. de Salazar, Benjamin J. Cowling, Marc Lipsitch, and Gabriel M. Leung. 2020. "Estimating Clinical Severity of COVID-19 from the Transmission Dynamics in Wuhan, China." *Nature Medicine* 26 (4): 506–10.

Zhao, Juanjuan, Jr., Quan Yuan, Haiyan Wang, Wei Liu, Xuejiao Liao, Yingying Su, Xin Wang, et al. 2020. "Antibody Responses to SARS-CoV-2 in Patients of Novel Coronavirus Disease 2019." *Infectious Diseases (except HIV/AIDS)*. medRxiv. https://doi.org/10.1101/2020.03.02.20030189.

Zou, Lirong, Feng Ruan, Mingxing Huang, Lijun Liang, Huitao Huang, Zhongsi Hong, Jianxiang Yu, et al. 2020. "SARS-CoV-2 Viral Load in Upper Respiratory Specimens of Infected Patients." *The New England Journal of Medicine* 382 (12): 1177–79.